\documentclass{ws-ijmpa}
\usepackage[super,compress]{cite}
\usepackage{graphicx}
\usepackage{graphicx}
\usepackage{multirow}
\usepackage{xcolor}
\usepackage{amsmath}
\usepackage{amsfonts}
\usepackage{amssymb}
\usepackage{url}
\usepackage{hyperref}
\usepackage{subfigure}
\usepackage{array}
\newcolumntype{M}[1]{>{\centering\arraybackslash}m{#1}}

\begin{document}
\markboth{Mayengbam Kishan Singh}{Modular $A_4$ symmetry in 3+1 active-sterile neutrino masses and mixings}

%
\catchline{}{}{}{}{}
%

\title{Modular $A_4$ symmetry in 3+1 active-sterile neutrino masses and mixings}

\author{Mayengbam Kishan Singh}

\address{Department of Physics, D.M. College of Science, Dhanamanjuri University.\\
Imphal, Manipur-795001, India \\
kishanmayengbam@gmail.com}

\author{S. Robertson Singh}

\address{Department of Physics, Modern College, Porompat.\\
Imphal, Manipur-795005, India\\
robsoram@gmail.com}

\author{N. Nimai Singh}

\address{Department of Physics, Manipur University, Canchipur.\\
Research Institute of Science and Technology (RIST),\\
Imphal, Manipur-795003, India\\
nimai03@yahoo.com}

\maketitle


\begin{abstract}
Motivated by the significance of modular symmetry in generating neutrino masses and flavor mixings, we apply the modular $A_4$ symmetry in a 3+1 scheme of active-sterile neutrino mixings. Neutrino oscillation observables in the 3$\sigma$ range are successfully reproduced through the vacuum expectation value of the modulus $\tau$. We also study phenomenologies related to the effective neutrino masses $m_{\beta}$ in tritium beta decay and $m_{\beta\beta}$ in neutrinoless double beta decay. Mixings between active neutrinos and eV scale sterile neutrino are analyzed in detail. The model also predicts the Dirac CP-violating phase $\delta_{CP}$ and Majonara phases $\alpha$ and $\beta$. The best-fit values of the neutrino mixing angles and ratios of two mass-squared differences are determined using minimum $\chi^2$ analysis. The best-fit values of the neutrino oscillation observables are predicted as $\sin^2\theta_{23}=0.573,$ $\sin^2\theta_{12}=0.300,$ $ \sin^2\theta_{13}=0.022$ and $r = 0.172$ for NH while $\sin^2\theta_{23}=0.550,$ $\sin^2\theta_{12}=0.303,$ $ \sin^2\theta_{13}=0.022$ and $r = 0.171$ for IH. We also observe that the predictions of effective neutrino mass parameters in the 3+1 scheme are significantly different from the three neutrino paradigm.

\end{abstract}

\ccode{PACS numbers:}


\section{Introduction}
The origin of neutrino masses and flavor structure is one of the most critical problems of particle physics's Standard Model (SM). Since the discovery of neutrino oscillation in various experiments such as SNO, SuperKamiokande, etc., various extensions of SM have been studied. Models with non-Abelian discrete symmetries such as $A_4$  \cite{altarelli2010discrete,king2007a4,babu2003underlying}, $S_4$  \cite{ma2006neutrino,altarelli2009revisiting,vien2022lepton,bazzocchi2013neutrino}, $S_3$ \cite{mohapatra2006s3,morisi2006flavor,Vien:2021uxw,grimus2005s3,koide2007s}, etc. find their distinct places in describing some of the experimental results as well as new predictions regarding unresolved problems of SM. Among these, the absolute mass scale of neutrino, Dirac CP-violating phase, Baryogenesis, Dark matter, etc., are some of the main questions in high-energy physics. 

Inspired by the observations of LSND  \cite{aguilar2001evidence} and MiniBooNE  \cite{aguilar2018significant,2021microboone}, many authors proposed the existence of a fourth state of neutrino called the sterile neutrino. Sterile neutrinos are incorporated along with the three neutrino theory in literatures in a 3+1 scheme  \cite{Zhang2011,giunti2020katrin,guintianrev, vien2022b,miranda2019revisiting,vien2021}, 3+1+1 scheme  \cite{kuflik2012neutrino,huang2013mev,fan2012light,nelson2011effects}, 3+2 scheme  \cite{donini2012minimal,archidiacono2012testing,babu2016light,karagiorgi2007leptonic} etc. One of the simplest extensions is the 3+1 scheme, where a singlet sterile neutrino is added to the three active neutrinos, and the sterile neutrino gets mass through a minimally extended seesaw mechanism (MES) \cite{Zhang2011}.
Many authors have used discrete symmetries to describe the neutrino masses and their flavor structures. The main disadvantages of such approaches are the inclusion of many hypothetical scalar fields and additional symmetry groups. Recently, an interesting approach has been proposed in which modular symmetry is used in which the non-Abelian discrete symmetries are its subgroups \cite{feruglio2019neutrino}. An important feature of the modular symmetry framework is that the Yukawa couplings can be transformed non-trivially as modular forms under the modular group. These modular forms are written as a function of a complex parameter $\tau$, called the modulus. Minimal or no scalar flavons are required to break the symmetry as the lepton masses are generated from the symmetry breaking by the VEV of the modulus $\tau$. Modular forms of level 3 denoted by $\Gamma(3)$ are isomorphic to $A_4$. Detailed analysis on modular $A_4$ groups and their application in neutrino model building is studied in Refs. \citen{feruglio2019neutrino,kobayashi2018neutrino,kobayashi2018modular}. There are other works based on numerous modular groups such as modular $A_4$  \cite{zhang2020modular,Beherascoto,mishra2022type,kobayashi2020type,okada2020radiative,
nomura2019modular,abbas2021fermion,nomurascoto,behera2022implications}, modular $A_5$  \cite{ding2019neutrino,novichkov2019modular,PhysRevD.103.095013,PhysRevD.103.076005,Behera2021eut}, modular $S_4$ \cite{wang2020minimal,penedo2019lepton,wang2021dirac,ding2019modular,
KobayasiS4,LiuS4}, etc., to study neutrino phenomenologies, dark matter, leptogenesis, etc. For instance, in Ref. \citen{abbas2021fermion},  modular $A_4$ is used to study lepton and quark masses and mixings based on inverse seesaw mechanism, Ref. \citen{behera2022implications} studies lepton mixings and leptogenesis in linear seesaw. In contrast, the scotogenic dark matter scenario is studied in Refs. \citen{Beherascoto,nomurascoto}. Other works in the literature study 3+1 active-sterile mixing using the conventional $A_4$ discrete symmetry group  \cite{das2019active,mksingh,vien2022b,Das3}. However,  modular $A_4$ symmetry has not been applied in 3+1 active-sterile neutrino mixing in MES mechanism. The main objectives and features of the present work are: (i) to develop a new model of neutrino masses and flavor mixings based on modular $A_4$ symmetry for 3+1 scheme based on MES mechanism, (ii) to establish a detailed analysis of active-sterile mixing for sterile neutrinos in the mass in eV scale, (iii) to reproduce low energy active neutrino mixing angles, mass-squared differences as well as Dirac and Majorana phases $\delta_{CP},\ \alpha$ and $\beta$ in the 3$\sigma$ ranges constrained by the cosmological Planck upper limit $\sum m_{i}<0.12$ eV, (iv) to predict the effective neutrino mass parameters $m_{\beta}$ and $m_{\beta\beta}$ from tritium beta decay and neutrinoless double beta decay $(0\nu\beta\beta)$ and (v) finally, to determine the best-fit values of the model parameters and neutrino observables using the minimum $\chi^2$ analysis.

As we shall discuss later, exciting results are obtained for the modulus $\tau$ in a narrow range within the fundamental domain. The model's predictability is improved by excluding hypothetical scalar triplet flavons compared to conventional model building. Significant results are obtained for the active-sterile mixing parameters $\vert U_{i4}\vert^2 $, (where $i=1,2,3$). Predictions in the phenomenological studies of effective neutrino masses $m_{\beta}$ and $m_{\beta\beta}$ reveal new results that future experiments could probe. For instance, we obtain distinctive results between the values of $m_{\beta}$ and $m_{\beta\beta}$ in three neutrinos (3$\nu$) mixings and 3+1 neutrino mixings. Future experiments could probe these results, and we may use them to validate or discard the existence of eV-scale sterile neutrinos. With hints of the non-existence of sterile neutrinos from the initial results of MicroBooNE, we shall study a phenomenological approach to the constraints of active-sterile neutrino mixings. The present paper is organized as follows. We present a detailed description of the model in section \ref{c5s2}, followed by the numerical analysis of the model in section \ref{section3}. Results of the numerical analysis are presented in section \ref{section4}. We conclude with a summary and discussion in section \ref{section6}.

\section{The structure of the model}\label{c5s2}
This model uses modular $A_4$ symmetry to study lepton masses and mixings in the 3+1 framework. We extend the SM particle contents with three right-handed neutrinos $N$ as triplet under modular $ A_4$ and a singlet sterile neutrino field $S$. The SM lepton doublet $L$ transforms as a triplet of $A_4$ while the Higgs represented as $H_u$ and $H_d$ having hypercharges $+1/2 $ and $-1/2$  are taken as singlets of $A_4$. They develop VEVs after the spontaneous symmetry breaking (SSB) with alignments along $ H_u = \left(0, v_u/\sqrt{2}\right)^T$ and $ H_d = \left( v_d/\sqrt{2}, 0\right)$ inducing fermion mass terms. The right-handed charged leptons $e_r, \mu_r$ and $\tau_r$ are assigned $1,\ 1^{\prime\prime},\ 1^{\prime}$ of modular $A_4$ respectively. As a result, there are three independent coupling constants $\alpha^{\prime},\ \beta^{\prime}$, and $\gamma^{\prime}$ in the Lagrangian of the charged lepton sector. We consider one scalar field $\zeta$, which is an $A_4$ singlet so that its vacuum expectation value (VEV) $v_{\zeta}$ fixes the scale of sterile neutrino mass matrix $M_S$. 
The MES mechanism has the advantage of naturally generating sterile neutrinos in eV or KeV scale based on the scale of $M_S$\footnote{ $M_S\sim \mathcal{O}(10^2)$GeV will give sterile neutrinos in eV scale. In contrast, $M_S\sim \mathcal{O}(10)$TeV will generate KeV scale sterile neutrino mass}. In modular $A_4$ symmetry, the Yukawa coupling transforms as an $A_4$ triplet modular function $Y =(y_1,y_2,y_3)^T$ having weight $k_Y$.  The complete particle contents of the model with their corresponding group charges and modular weights $k_i$ are given in Table \ref{Table1}. The components of modular form $Y$ of weight 2, which transforms as triplet under $A_4$ can be expressed in terms of the Dedekind eta-function $\eta(\tau)$ and its derivative $\eta^{\prime}(\tau)$ as \cite{feruglio2019neutrino}

\begin{eqnarray}
y_1(\tau)= & \frac{i}{2\pi}\left[\frac{\eta^{\prime}(\frac{\tau}{3})}{\eta (\frac{\tau}{3})}+ \frac{\eta^{\prime}(\frac{\tau+1}{3})}{\eta (\frac{\tau+1}{3})} + \frac{\eta^{\prime}(\frac{\tau+2}{3})}{\eta (\frac{\tau+2}{3})} - \frac{27\eta^{\prime}(3\tau)}{\eta(3\tau)}\right],\nonumber \\
y_2(\tau)= &\frac{-i}{\pi}\left[\frac{\eta^{\prime}(\frac{\tau}{3})}{\eta (\frac{\tau}{3})}+\omega^2 \frac{\eta^{\prime}(\frac{\tau+1}{3})}{\eta (\frac{\tau+1}{3})} +\omega \frac{\eta^{\prime}(\frac{\tau+2}{3})}{\eta (\frac{\tau+2}{3})} \right],\nonumber\\
y_3(\tau)= &\frac{-i}{\pi}\left[\frac{\eta^{\prime}(\frac{\tau}{3})}{\eta (\frac{\tau}{3})}+\omega \frac{\eta^{\prime}(\frac{\tau+1}{3})}{\eta (\frac{\tau+1}{3})} +\omega^2 \frac{\eta^{\prime}(\frac{\tau+2}{3})}{\eta (\frac{\tau+2}{3})} \right].\label{C5etafn}
\end{eqnarray}

where $\omega = e^{2\pi i/3}$ and $\eta(\tau)$ is defined as 
\begin{equation}
\eta(\tau)=q^{1/24}\prod_{n=1}^{\infty}(1-q^n),  \ \ \ \ q \equiv e^{i 2\pi \tau}.
\end{equation}
The overall coefficients in (\ref{C5etafn}) is one possible choice; it cannot be uniquely determined. The $q-$expansions of the Yukawa components are expressed as 
\begin{eqnarray}
y_1(\tau)& = 1+12q+36q^2+12q^3+... \nonumber\\
y_2(\tau)& = -16q^{1/3}(1+7q+8q^2+...)\\
y_3(\tau)& = -18q^{2/3}(1+2q+5q^2+...).\nonumber
\end{eqnarray}
In a modular symmetry framework, an interaction is invariant under the symmetry when the sum of the modular weights for each associated field and coupling is zero, and it is also invariant under the discrete symmetry (such as $A_4$ symmetry).

\begin{table}
\centering
\tbl{Particle contents of the model and their group representations and hypercharges.}
{\begin{tabular}{@{}m{1.5cm}m{1.3cm}m{1.4cm} m{1.5cm} m{0.5cm}m{0.5cm} m{0.5cm}@{}} \toprule
$Fields$ & $L$ & $e_r^c, \mu_r^c,\tau_r^c$ & $H_{u,d}$ & $N^c$ & $S^c$ & $\zeta$   \\
&  &  \\ \colrule
$SU(2)_L$& 2 & 1 & 2 & 1 & 1  &1   \\
$U(1)_Y$ & $-1/2$ & $1$& $\pm$ 1/2 & 0& 0&0 \\
$A_4$& 3 &1,1$^{\prime\prime}$,1$^{\prime}$ &1 & 3 & 1 &1   \\
$k_i$&$k_L$ &$k_e$  &$k_H$ &$k_N$ & $k_S$ &$k_{\zeta}$ \\
\hline
\end{tabular}\label{Table1}}
\end{table}

The Lagrangian of charged lepton sector invariant under $A_4$ symmetry is given below.
\begin{eqnarray}
-\mathcal{L}_{c} =\ \alpha^{\prime} e_r^c( H_d L Y)_1 + \beta^{\prime} \mu_r^c (H_d L Y)_{1^{\prime}} + \gamma^{\prime}\tau_r^c(H_d L Y)_{1^{\prime\prime}} + h.c.
\label{C5chargelepton}
\end{eqnarray}
For the neutrino sector, the Lagrangian is 
\begin{eqnarray}\label{C5md}
-\mathcal{L}_{D} &=&\ g_i (N^c)^TH_uLY + h.c \\ \nonumber
  & = &\ g_1 (N^c)^TH_u(LY)_{3S} + g_2 (N^c)^TH_u(LY)_{3A} + h.c \\
-\mathcal{L}_{R} &=& \ \lambda Y (N^c)^T N + h.c \\
-\mathcal{L}_s &=& \ \zeta Y S^cN +h.c
\label{C5MR}
\end{eqnarray}
where, $\alpha^{\prime}, \beta^{\prime}$,$\gamma^{\prime},\ g_1,\ g_2,$ and $\lambda$ are constant coefficients. The subscripts $`3S$' and $`3A$' denote the symmetric and antisymmetric product of two triplet fields in $A_4$ symmetry. The above interaction terms for the charged lepton and neutrino sector are invariant under modular $A_4$ symmetry if the weights $k_i$ satisfy the following conditions,
\begin{eqnarray}
k_e + k_H+k_L =0,\ \ 
k_H+k_L+k_N+k_Y=0, \nonumber \\  
k_N+k_N+k_Y=0, \ \ \ \ \
k_{\zeta}+k_S+k_N+k_Y=0.
\end{eqnarray}
For modular form of weight $k_Y=2,$ we have used the following assignments of modular weights,
\begin{equation}
k_N=k_L=-1, \ k_H=0,k_S=1,\ k_{\zeta}=-2, \ \mbox{and} \ k_e=-1.
\end{equation}
Other higher dimensional terms such as $\frac{g^{\prime}}{\Lambda}H_u\zeta S^c(LY)_1$, $\frac{\lambda_1}{\Lambda^2}YYY(N_i^c)^TN_i\zeta\zeta$, $\frac{\delta_1}{\Lambda^2}YYNS\zeta\zeta\zeta$, $\frac{\delta_2}{\Lambda^3}YYSS\zeta\zeta\zeta\zeta$, etc. in the neutrino sector are also allowed by the symmetry of the model. However, their contributions are negligibly small and have been ignored in our analysis. 

After the electroweak symmetry breaking, using (\ref{C5md}) - (\ref{C5MR}), the Dirac, Majorana, and sterile neutrino mass matrices are given as

\begin{eqnarray}
M_D &= &\ v_u \left(
\begin{array}{ccc}
 2 g_1 y_1 & y_3 (g_2-g_1) & y_2 (-g_1-g_2) \\
 y_3 (-g_1-g_2) & 2 g_1 y_2 & y_1 (g_2-g_1) \\
 y_2 (g_2-g_1) & y_1(-g_1-g_2) & 2 g_1 y_3 \\
\end{array}
\right); \\
M_R &=&\ \lambda\left(
\begin{array}{ccc}
 2y_1 & -y_3 & -y_2 \\
 -y_3 & 2 y_2 & -y_1 \\
 -y_2 & -y_1 & 2 y_3 \\
\end{array}
\right);\\
M_s &=&\ v_{\zeta}\left(
\begin{array}{ccc}
y_1 & y_3 & y_2 \\
\end{array}
\right).
\label{matrix}
\end{eqnarray}
Whereas, from (\ref{C5chargelepton}), the charged lepton mass matrix becomes 
\begin{eqnarray}
m_L = v_d \left(\begin{array}{ccc}
\alpha^{\prime} & 0 & 0 \\ 
0 & \beta^{\prime} & 0 \\ 
0 & 0 & \gamma^{\prime}
\end{array} \right)\times 
\left(\begin{array}{ccc}
y_1 & y_3 & y_2 \\ 
y_2 & y_1 & y_3 \\ 
y_3 & y_2 & y_1
\end{array} \right).
\end{eqnarray}
 It is convenient to diagonalise the $m_L$ using a hermitian matrix $M_{L}=m^{\dagger}_L m_L$ as
\begin{eqnarray}
M^{diag}_L = \mathcal{U}^{\dagger}_L M_L \mathcal{U}_L.
\label{mldiag}
\end{eqnarray}

In the MES mechanism, we obtain a leading order of the active neutrino mass matrix $m_{\nu}$ as well as the sterile mass $m_s$ given as follows.
\begin{equation}
m_{\nu} \simeq M_DM_R^{-1}M_S^T\left(M_S M_R^{-1}M_S^T\right)^{-1}M_S\left(M_R^{-1}\right)^T M_D^T-M_DM_R^{-1}M_D^T;
\label{C5mv}
\end{equation}
\begin{equation}
m_s \simeq - M_SM_R^{-1}M_S^T.
\label{C5ms}
\end{equation}

The full $(4\times 4)$ active-sterile neutrino mass matrix is diagonalised by a unitary $(4\times 4)$ mixing matrix given by  \cite{v441982}
\begin{equation}
V \simeq \left(\begin{array}{ccc}
(1-\frac{1}{2}RR^{\dagger})U & R \\ 
-R^{\dagger}U & 1-\frac{1}{2}R^{\dagger}R
\end{array} \right),
\label{C5V44}
\end{equation}
where $U$ represents the $3\times 3$ active neutrino mixing matrix and $R$ represents the strength of active-sterile mixing given by
\begin{eqnarray}
R=&M_DM_R^{-1}M_S^T(M_SM_R^{-1}M_S^T)^{-1}.
\label{C5R}
\end{eqnarray}
Using (\ref{C5ms}) and (\ref{C5R}), the sterile neutrino mass and active-sterile mixing strength are given by 
 \begin{eqnarray}
 m_s &=&\ \frac{v_\zeta^2\ y_1 (y_1^3 - 12 y_1 y_2 y_3 - 8 (y_2^3 + y_3^3))}{2\lambda (y_1^3 + y_2^3 - 3 y_1 y_2 y_3 + y_3^3)}
 \end{eqnarray}
 \begin{eqnarray*}
 R =\ f \left(\begin{array}{c}
 y_1^3 + (1 + 3 g) y_2^3 - 3 y_1 y_2 y_3 + (1 - 3 g) y_3^3 \\ 
 y_3 (y_1^3 + y_2^3 - 3 y_1 y_2 y_3 + y_3^3) + 
     g (-3 y_1^2 y_2^2 - 2 y_1^3 y_3 + y_2^3 y_3 + 3 y_1 y_2 y_3^2 + 
        y_3^4) \\ 
 (y_1^3 (y_2 + 2 g y_2) - 
     3 (1 + g) y_1 y_2^2 y_3 + 
     3 g y_1^2 y_3^2 - (-1 + g) y_2 (y_2^3 + y_3^3))\end{array} \right)
 \end{eqnarray*}
where \begin{equation}
f = \frac{2 g_1 v}{
  v_{\zeta}\left[ y_1^3 - 12 y_1 y_2 y_3 - 
     8 (y_2^3 + y_3^3)\right]}.
\end{equation}   

Deviation of $U$ from unitarity due to the presence of sterile neutrino is determined by $1- \frac{1}{2}RR^{\dagger}$. The $4\times 4$ neutrino mixing matrix $V$ can be parametrized by six mixing angles $(\theta_{12},\ \theta_{13},\ \theta_{23},\ \theta_{14},\theta_{24},\ \theta_{34})$, three Dirac phases $(\delta_{CP},\ \delta_{14},\ \delta_{24})$ and three Majorana phases $(\alpha,\ \beta,\ \gamma)$ \cite{gariazzo2016light}. However, in the active neutrino sector, we note that the lightest neutrino mass in the MES mechanism is vanishing, i.e., $m_1=0(m_3=0)$ for NH(IH). Thus, we are left with only one Majorana CP-violating phase $\beta$ while $\alpha$ will be vanishing (Here, $\gamma$ can be re-phased out)\cite{nath2016understanding}. This result will be evident from the numerical analysis as well.
 
In the above parametrization, the unitary matrix V can be written as,
\begin{equation}
V = O_{34}\tilde{O}_{24}\tilde{O}_{14}O_{23}\tilde{O}_{13}O_{12}.P
\label{V44p}
\end{equation}
where $O_{ij}$ are rotation matrices and P is a diagonal phase matrix given by 
\begin{equation}
P = diag\{e^{i\alpha/2},e^{i (\beta + \delta_{13})/2},e^{i(\gamma +\delta_{14})/2},1\}.
\end{equation}

The six neutrino mixing angles are determined from the mixing elements of $V$ using the following relations 

\begin{eqnarray} \label{C1anglessolve}
\sin^2\theta_{14}\ =\ \vert V_{e4}\vert ^2,\ \  
\sin^2\theta_{24}\ =\ \frac{\vert V_{\mu 4}\vert ^2}{1-\vert V_{e4}\vert ^2},\ \ 
\sin^2\theta_{34}\ =\ \frac{\vert V_{\tau 4}\vert ^2}{1-\vert V_{e4}\vert ^2-\vert V_{\mu 4}\vert ^2},\nonumber \\
\sin^2\theta_{12}\ =\ \frac{\vert V_{e2}\vert ^2}{1-\vert V_{e4}\vert ^2-\vert V_{e3}\vert ^2}, \ \ 
\sin^2\theta_{13}\ =\ \frac{\vert V_{e3}\vert ^2}{1-\vert V_{e4}\vert ^2},
\end{eqnarray}
\begin{eqnarray}
 \sin^2\theta_{23}\ =&\ \frac{\vert V_{e3}\vert ^2(1-\vert V_{e4}\vert ^2)-\vert V_{e4}\vert ^2\vert V_{\mu 4}\vert ^2}{1-\vert V_{e4}\vert ^2-\vert V_{\mu 4}\vert ^2}+ \frac{\vert V_{e1}V_{\mu 1}+V_{e2}V_{\mu 2}\vert ^2(1-\vert V_{e4}\vert ^2)}{(1-\vert V_{e4}\vert ^2-\vert V_{e3}\vert ^2)(1-\vert V_{e4}\vert ^2-\vert V_{\mu 4}\vert ^2)}.\nonumber 
\end{eqnarray}
where $V_{ij}$ are the elements of mixing matrix in (\ref{V44p}).

The Dirac CP-violating phase $\delta_{CP}$, is related to the $4\times 4$ mixing matrix $V$ through the Jarlskog invariant $J.$ According to the parametrisation given in (\ref{V44p}), the Jarlskog invariant defined as $J = Im[V_{e1}V_{\mu 2}V^*_{e2}V^*_{\mu 1}]$  takes the form  \cite{KUMAR2020115082}
\begin{equation}
J = J_3^{cp} c_{14}^2c_{24}^2 + s_{24}s_{14}c_{24}c_{23}c^2_{14}c^3_{13}c_{12}s_{12}\sin(\delta_{14}-\delta_{24}),
\label{C5deltasolve}
\end{equation}
where $J_3^{cp} = s_{23}c_{23}s_{12}c_{12}s_{13}c_{13}^2 \sin \delta_{CP}$ is the Jarlskog invariant in the three neutrino framework and $s_{ij} = \sin\theta_{ij},c_{ij}=\cos\theta_{ij}$ are the neutrino mixing angles. Similarly, the two physical Majorana phases $\alpha$ and $\beta$ are determined from $V$ using the invariants $I_1$ and $I_2$ defined as follows 
\begin{eqnarray}\label{C5majoranaphase1}
I_1 = Im[V_{e1}^*V_{e2}]\ =\ c_{12} c_{13}^2 c_{14}^2 s_{12} \sin\left(\frac{\alpha}{2}\right), \\ 
I_2 = Im[V_{e1}^*V_{e3}]\ =\ c_{12}c_{13}c_{14}^2s_{13}\sin\left(\frac{\beta}{2}-\delta_{CP}\right).
\label{C5majoranaphase2}
\end{eqnarray}

Other important parameters in neutrino physics are the effective neutrino mass $m_{\beta\beta}$ and effective electron neutrino mass $m_{\beta}$. A combined analysis from KamLAND-Zen \cite{Kamland} and GERDA provided an upper bound on $m_{\beta\beta}$ in the range $m_{\beta\beta} < (0.071-0.161$) eV \cite{agostini2018improved,goswami}. Recent results from the KATRIN-2020 \cite{aker2020first} experiment constrains the effective electron neutrino mass $m_{\beta}$ to be less than 1.1 eV. However, this upper bound has been updated to $m_{\beta}<0.8$ eV in the latest results of KATRIN-2022  \cite{katrin2022direct}. These parameters are determined from the neutrinoless double beta decay and tritium beta decay respectively, using the relations \cite{Hagstot}

\begin{eqnarray}
m_{\beta\beta}=\vert\sum_{j=1}^4 V_{ej}^2 m_j\vert
\label{mbbeq}
\end{eqnarray}
\begin{equation}
m_{\beta} = \left(\sum_{i=1}^4\vert V_{ei}^2\vert m_{i}^2\right)^{1/2}.
\label{mbeq}
\end{equation}
The presence of eV scale sterile neutrino mixings with active neutrinos have significant affects on the predictions of $m_{\beta\beta}$ and $m_{\beta}$ as pointed out in Ref. \citen{giunti2015predictions}.

\section{Numerical analysis}\label{section3}
In this section, we describe the steps of the detailed numerical analysis of the model. The charged lepton mass matrix $m_L$ depends on the free parameters $\alpha^{\prime}, \beta^{\prime}$, $\gamma^{\prime} $ and $\tau.$ We define the complex parameter $\tau$ as 
\begin{equation}
\tau=Re[\tau] + i\ Im[\tau].
\end{equation} 
The fundamental domain of $\tau$ is given in Ref.  \citen{feruglio2019neutrino}. Without the loss of generality,  $\alpha^{\prime},\ \beta^{\prime}$ and $\gamma^{\prime} $ can be taken to be real and positive. Given the complex parameter $\tau,$ we can solve these unknown coefficients using the following three identities 
\begin{eqnarray}
Tr[M_L] = m_e^2 + m_{\mu}^2 + m_{\tau}^2, \\
Det[M_L] = m_e^2 \times m_{\mu}^2 \times m_{\tau}^2, \\
Tr[M_L]^2/2 - Tr[M_L^2]/2 = m_em_{\mu} + m_{\mu}m_{\tau} + m_{\tau}m_e.
\end{eqnarray}
where $m_e , m_{\mu}$ and $ m_{\tau}$ are the charged lepton masses.

\begin{table}[ph]
\tbl{Updated global-fit data for three neutrino oscillation, NuFIT 5.3, 2024 \cite{nufit}. For 3+1 mixing, data is taken from  Refs.  \citen{barrylight,vien2022b,Gariazzo_2016}.}
{\begin{tabular} {@{}ccc @{}}
 \toprule
\rule{0pt}{4ex} Parameter &	Normal Hierarchy (best-fit$\pm 1\sigma$) &	Inverted Hierarchy (best-fit$\pm 1\sigma$)  \\
& & \\
\hline
\rule{0pt}{4ex} $\vert\Delta m^2_{21}\vert: [10^{-5} eV^2]$ & 6.82 – 8.03 $(7.41^{+0.21}_{-0.20})$  &	 6.82 – 8.03 $(7.41^{+0.21}_{-0.20})$ \\
$\vert\Delta m^2_{31}\vert: [10^{-3} eV^2]$	& 2.428 – 2.597 $(2.511^{+0.028}_{-0.027})$  & 2.408 – 2.581 $(2.498^{+0.032}_{-0.025})$ \\

$\sin^2\theta_{12} $	& 0.270 – 0.341 $(0.303^{+0.012}_{-0.011})$ &  0.270 – 0.341 $(0.303^{+0.012}_{-0.011})$  \\
$\sin^2\theta_{23}$ &0.406 – 0.620	$(0.572^{+0.018}_{-0.023})$  &0.412 – 0.623 $(0.578^{+0.016}_{-0.021})$ 		 \\
$\sin^2\theta_{13}/10^{-2}$ & 2.029 – 2.391	$(2.203^{+0.056}_{-0.059})$ & 2.047 – 2.396 $(2.219^{+0.060}_{-0.057})$  \\
$\delta_{\rm CP}/^o$ &	108 - 404 $(197^{+42}_{-0.25})$	& 192 - 360 $(286^{+27}_{-32})$	 \\
$r=\sqrt{\frac{\Delta m_{21}^2}{\vert\Delta m_{3l}^2\vert}} $ & 0.1675 - 0.1759 (0.1718)  & 0.1683 - 0.1765 (0.1722)\\
$\vert U_{14}\vert^2 $  & 0.012 - 0.047 & 0.012 - 0.047 \\
$\vert U_{24}\vert^2 $  & 0.005 - 0.03 &  0.005 - 0.03 \\
$\vert U_{34}\vert^2 $  & 0 - 0.16 & 0 - 0.16 \\
& \\
\hline
\end{tabular}\label{data}}
\end{table} 

We randomly scan the real and imaginary parts of $\tau$ in the positive half of the fundamental domain, 
\begin{equation}
 Re[\tau] = [0,0.5],\ Im[\tau]= [0.6,2].
\label{pspace}
\end{equation} 

Using $v_d = 246$ GeV and $m_e = 0.51099$ MeV, $m_{\mu} = 105.65837$ MeV and $m_{\tau}= 1776.86$ MeV, taken from PDG \cite{pdg2022}, the values of $\alpha^{\prime},\ \beta^{\prime}$ and $\gamma^{\prime} $ are solved using the above identities. Once the charged lepton mass matrix is completely determined, it can be numerically diagonalized using (\ref{mldiag}).

\begin{figure}
\centering
\subfigure[]{
    \includegraphics[width=0.46\textwidth]{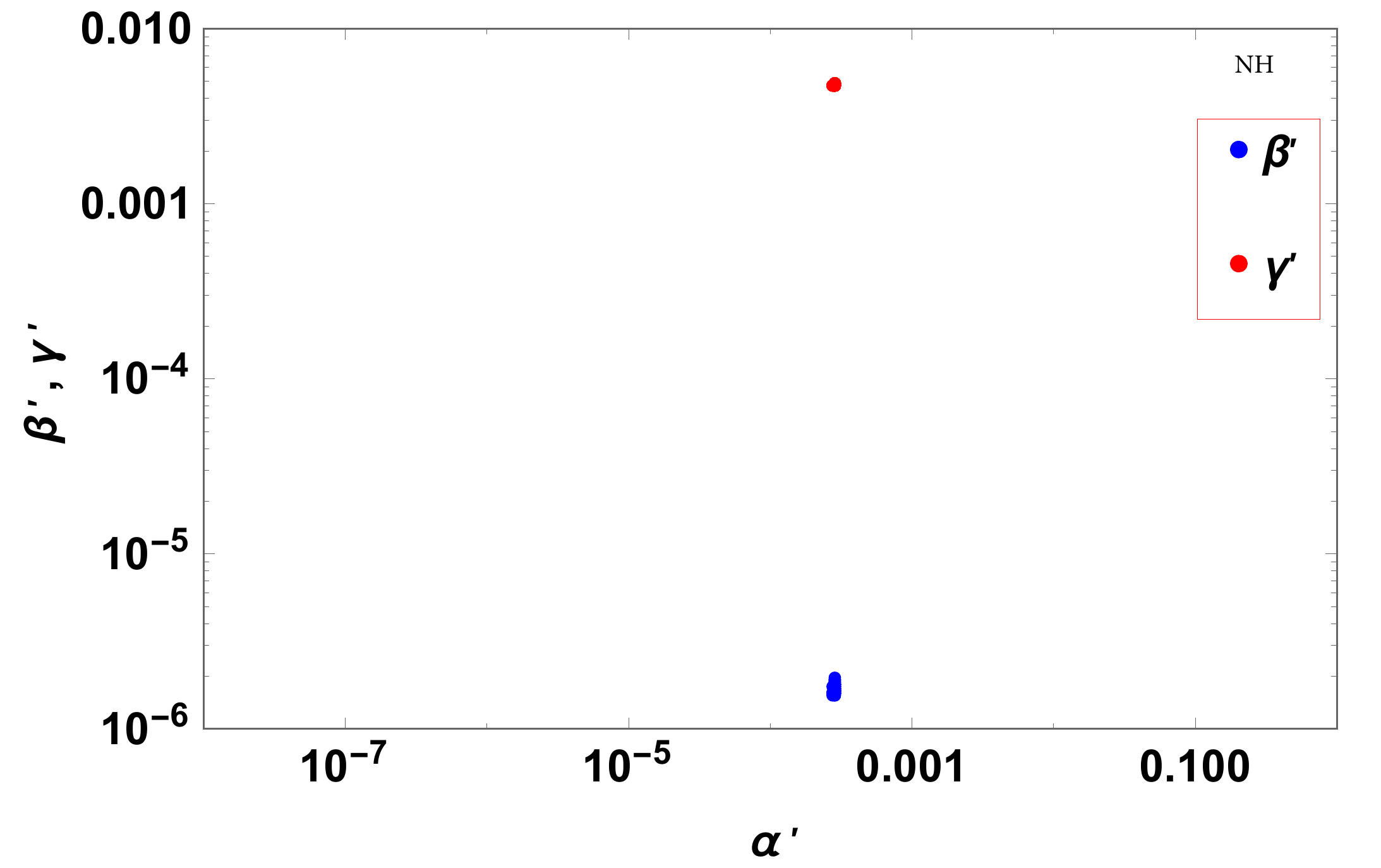}}
    \label{chargelepton}
 \quad
 \subfigure[]{
    \includegraphics[width=0.42\textwidth]{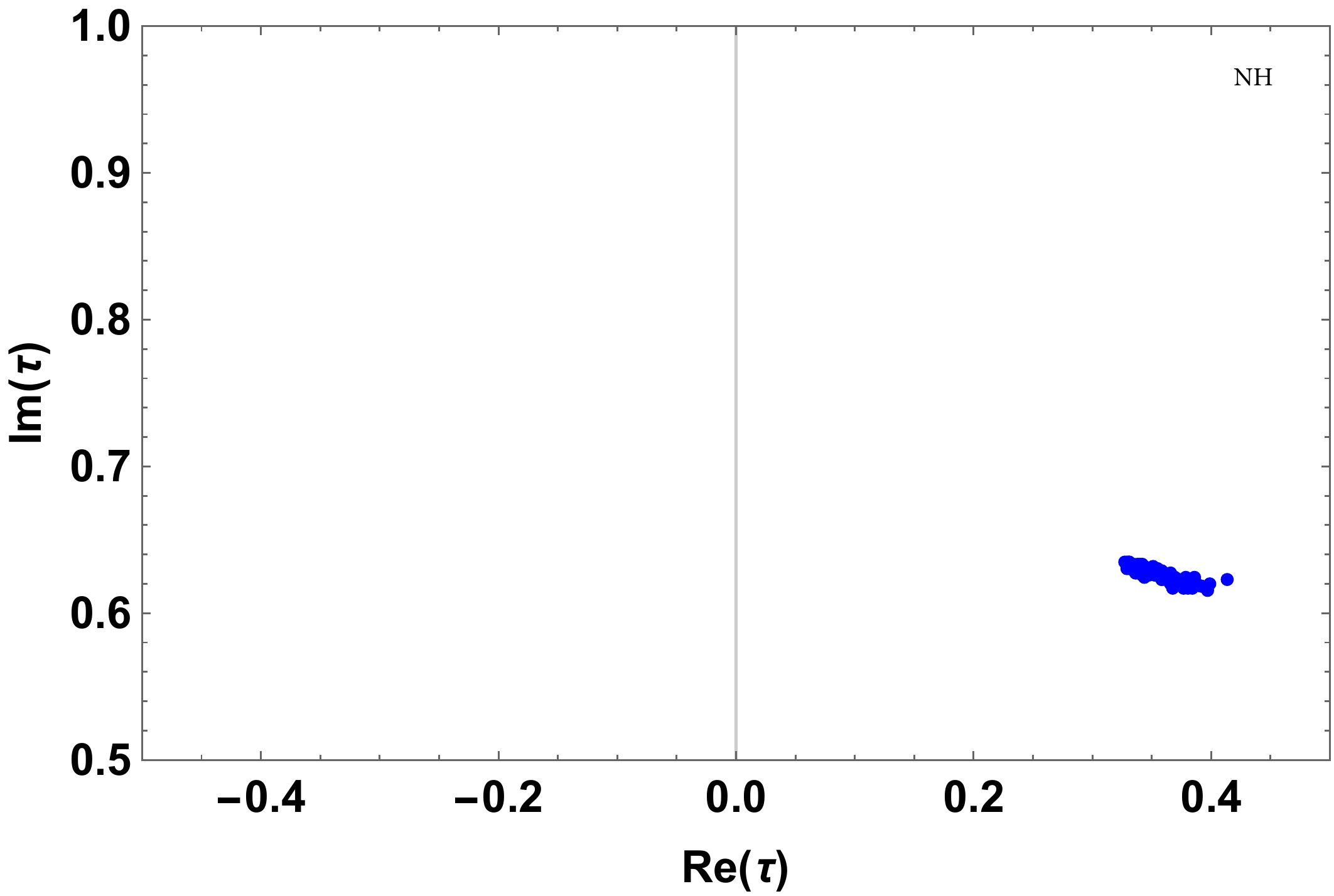}}
    \label{tauplot}
    \quad
  %
  \caption{\footnotesize (a) Variations of charged lepton Yukawa coefficients among themselves. (b)  Plot of allowed regions of $\tau$ constrained by neutrino oscillation data for normal hierarchy. }
  \label{C5tauchargeplot}
\end{figure}

\begin{figure}
\centering
\subfigure[]{
    \includegraphics[width=0.46\textwidth]{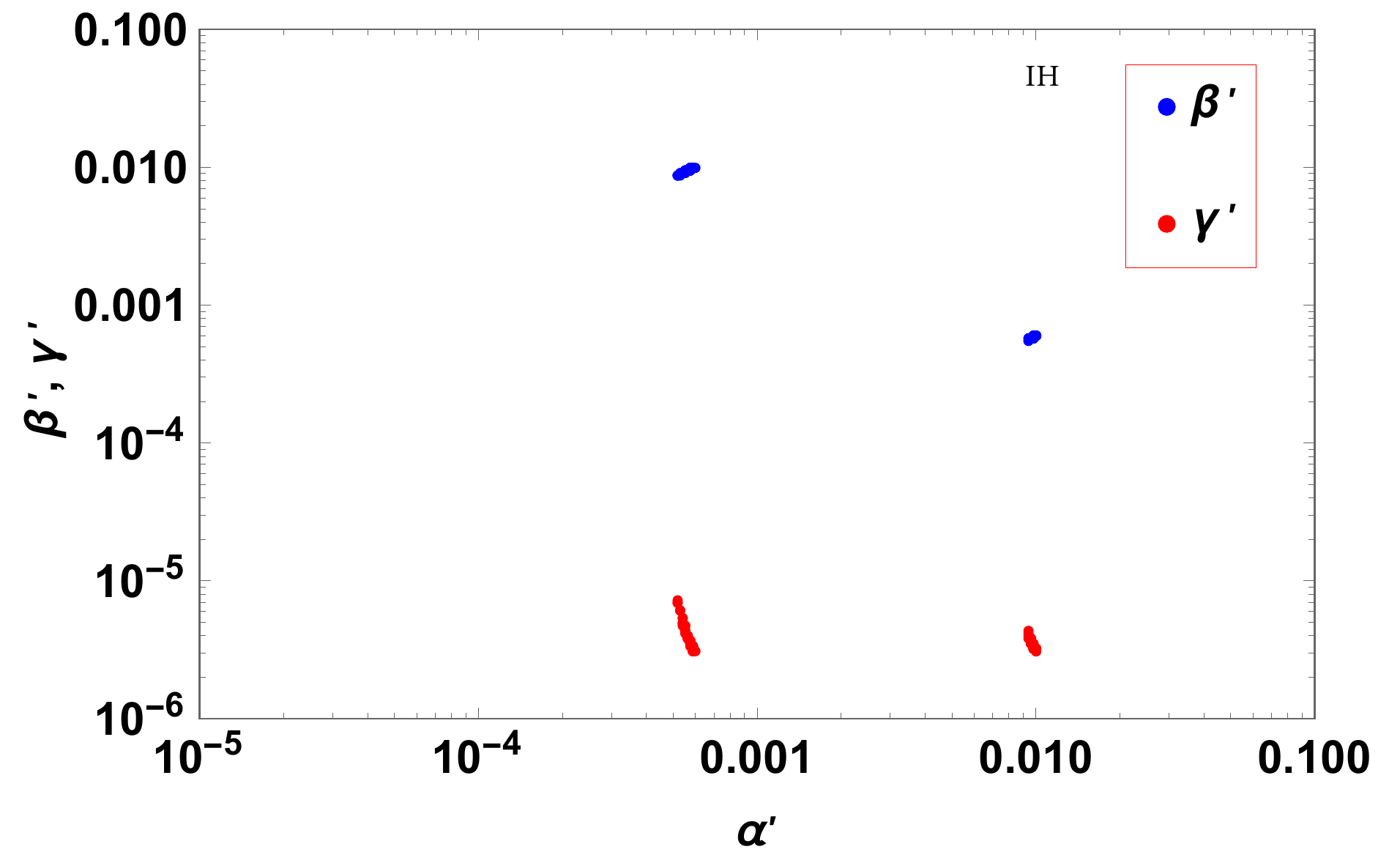}}
    \label{chargeleptonIH}
 \quad
 \subfigure[]{
    \includegraphics[width=0.42\textwidth]{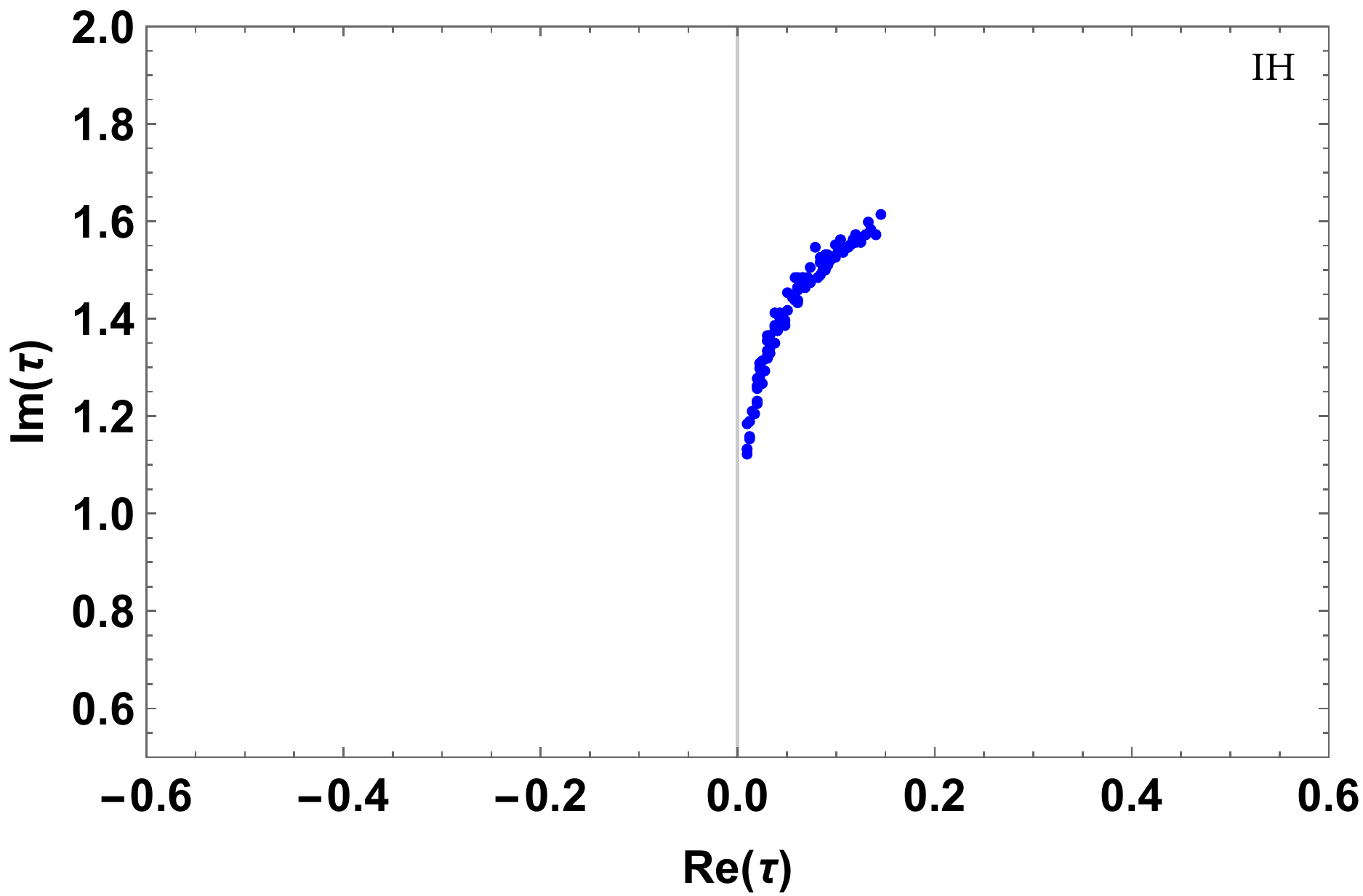}}
    \label{tauplotIH}
    \quad
  %
  \caption{\footnotesize (a) Variations of charged lepton Yukawa coefficients among themselves. (b)  Plot of allowed regions of $\tau$ constrained by neutrino oscillation data for inverted hierarchy. }
  \label{C5tauchargeplotIH}
\end{figure}

The active neutrino mass matrix given in (\ref{C5mv}) for the neutrino sector depends only on two complex parameters $g\equiv g_2/g_1$ and $\tau$,  up to an overall factor $v_u^2g_1^2/\lambda$. The modular symmetry is broken on fixing $\tau$, and the neutrino mass eigenvalues, three mixing angles, and the Dirac and Majorana phases are completely determined. The model becomes very predictive since the number of free parameters is less than the number of neutrino observables. The absolute scale of active neutrino masses is fixed by adjusting the overall factor $v_u^2g_1^2/\lambda$. With the values of $\tau$ obtained from the charged lepton sector, we give the following values as input parameters, 
\begin{equation}
 \vert g\vert = [0,10], \ \ \lambda = 10^{14}\ \mbox{GeV}, \ \ v_{\zeta} = [10,100]\ \mbox{GeV}\ \  \mbox{and}\ \ \phi_g = [-\pi,\pi].
\label{inputs}
\end{equation} 

Similar to the charged lepton sector, the active neutrino mass matrix $m_{\nu}$ is numerically diagonalized using the relation 
\begin{eqnarray}
 M_{\nu}^{diag} = \mathcal{U}^{\dagger}_{\nu} M_{\nu} \mathcal{U}_{\nu}
\end{eqnarray}
where $M_{\nu} = m_{\nu}^{\dagger}m_{\nu}$ is a hermitian matrix and $M_{\nu}^{diag} = diag(m_1^2,m_2^2,m_3^2)$. Here, the active neutrino mass eigenvalues are $m_1,\ m_2$, and $m_3$. Thus, the $(3\times 3)$ active neutrino PMNS mixing matrix will be given by 
\begin{eqnarray}
U = \mathcal{U}^{\dagger}_{L}\mathcal{U}_{\nu}.
\end{eqnarray}

We filter the values of the model parameters using the 3$\sigma$ bounds of the three mixing angles $\sin^2\theta_{12},\sin^2\theta_{13},\sin^2\theta_{23}$ and the ratio of neutrino mass squared differences $r= \sqrt{\Delta m_{21}^2/\Delta m_{31}^2}= m_2/m_3$ for normal hierarchy (NH) $(m_1 \approx 0\ll m_2 < m_3\ll m_4)$ and $ r=\sqrt{\Delta m_{21}^2 /\vert\Delta m_{32}^2\vert}= \sqrt{1-m_1^2/m_2^2}$ for inverted hierarchy (IH) $(m_3\approx0\ll m_2<m_1 \ll m_4)$ given in table \ref{data}. The sum of the active neutrino masses $\sum m_i = m_1 + m_2 + m_3 $ is also constrained by the Cosmological Planck upper limit $\sum m_i <0.12$ eV.

The allowed solutions for the charged lepton Yukawa coefficients $\alpha^{\prime},\ \beta^{\prime}$ and $\gamma^{\prime}$ are shown as correlation plots among themselves in figure \ref{C5tauchargeplot}(a) for NH and figure \ref{C5tauchargeplotIH}(a) for IH. In the fundamental domain of $\tau$, the model predicts both the normal hierarchy (NH) and inverted hierarchy (IH) of neutrino masses separately in two distinct regions.

\begin{figure}
\centering
\subfigure[]{
    \includegraphics[width=0.45\textwidth]{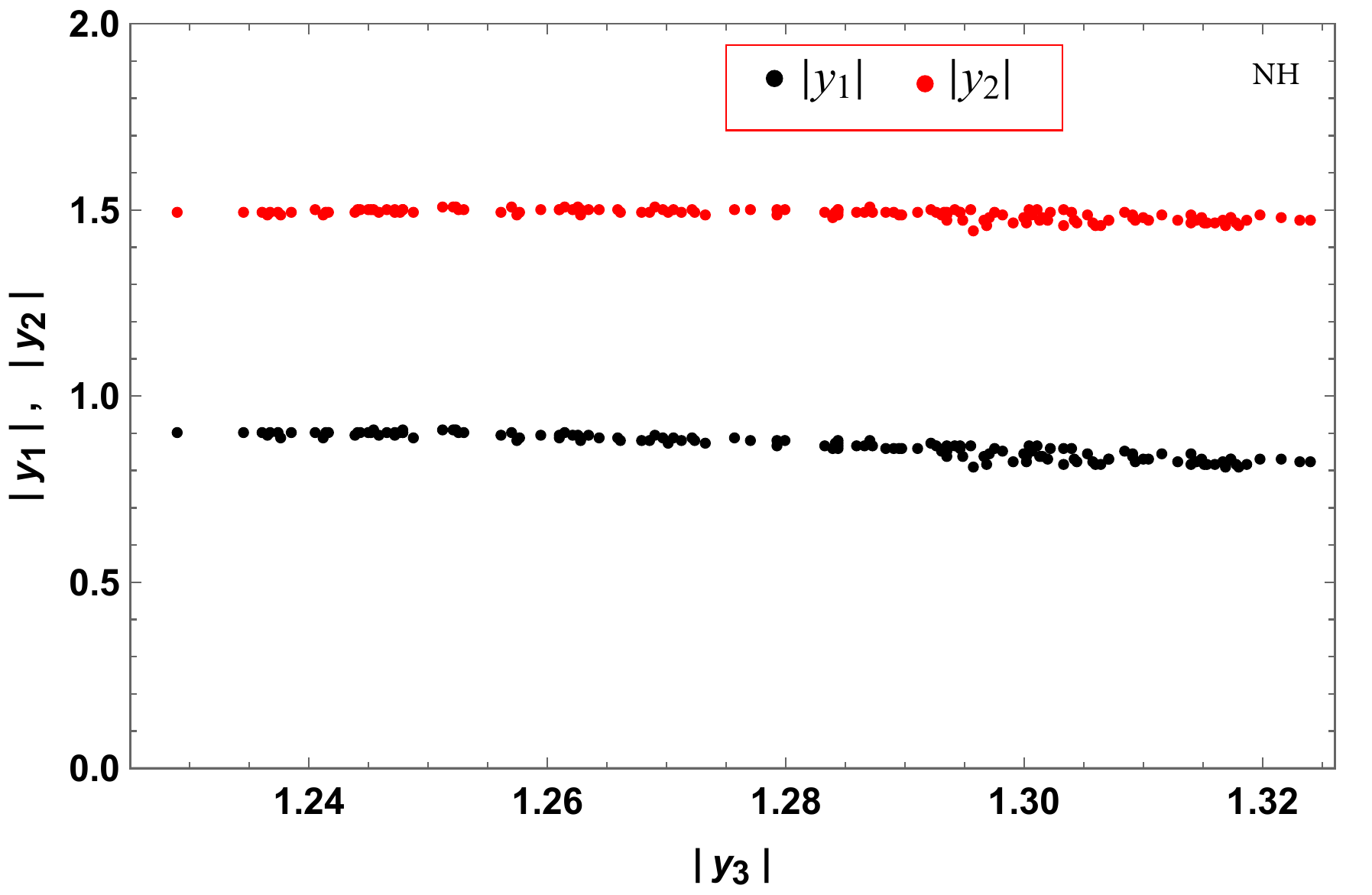}}
    \label{yukawa}
 \quad
\subfigure[]{
    \includegraphics[width=0.45\textwidth]{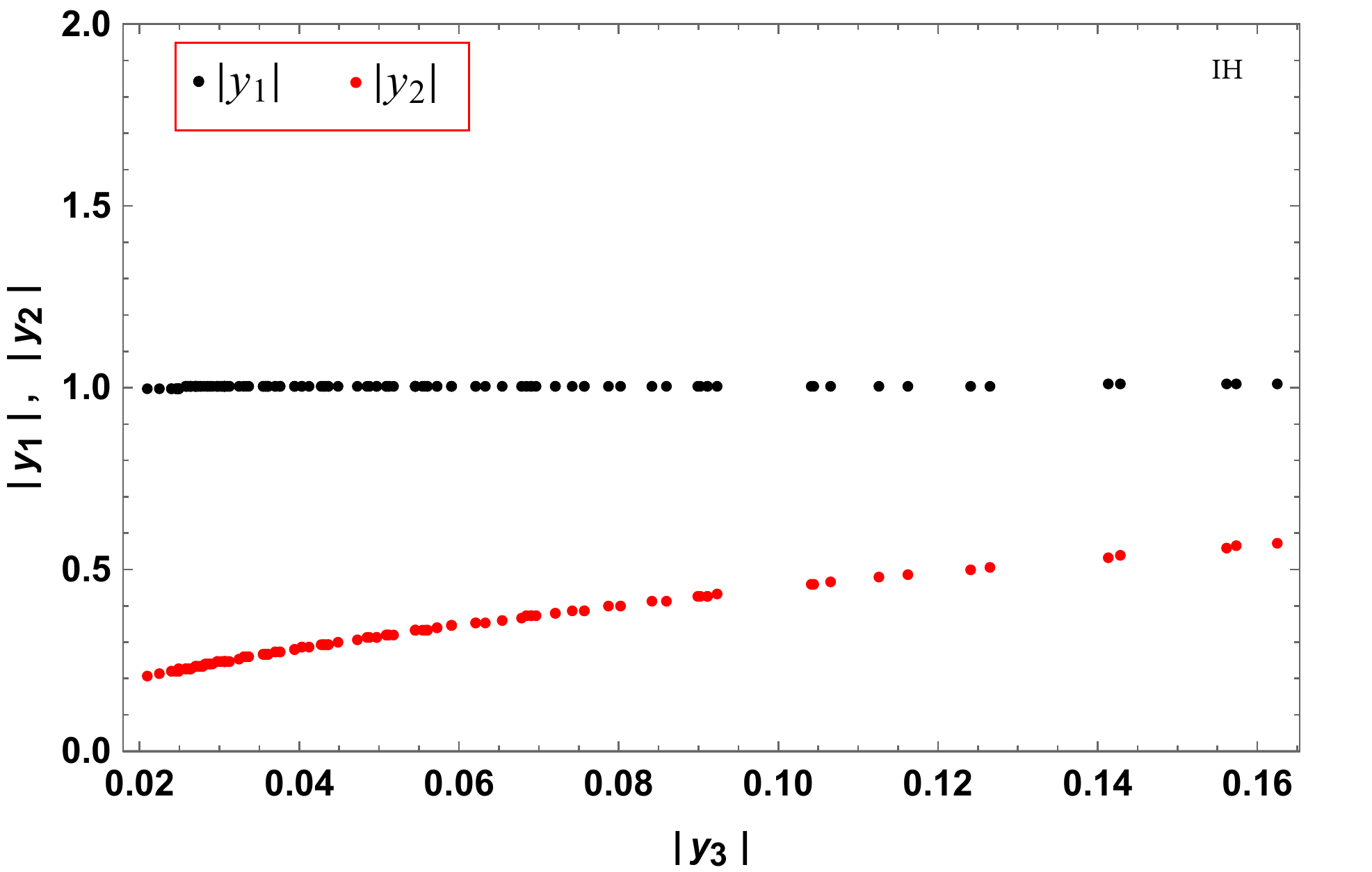}}
    \label{yukawaIH}
    \quad
  %
  \caption{\footnotesize Figure shows the  variation of Yukawa components $\vert y_1\vert,\vert y_2\vert$ with $\vert y_3\vert$ for NH in (a) and IH in (b).  }
  \label{yukawaplot}
\end{figure}
\begin{figure}
\centering
\subfigure[]{
    \includegraphics[width=0.45\textwidth]{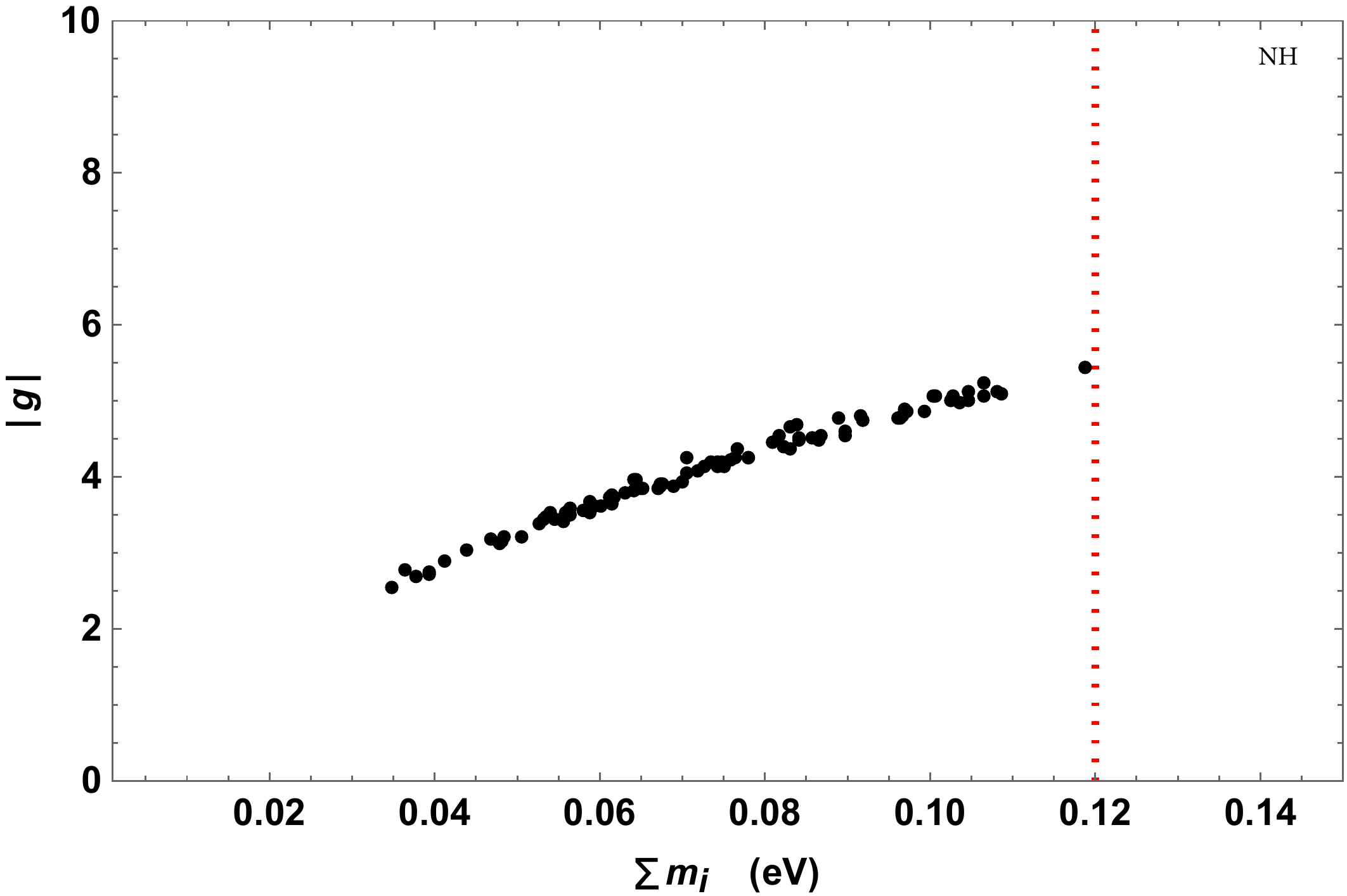}}
    \label{gvssum}
  \quad
\subfigure[]{
    \includegraphics[width=0.45\textwidth]{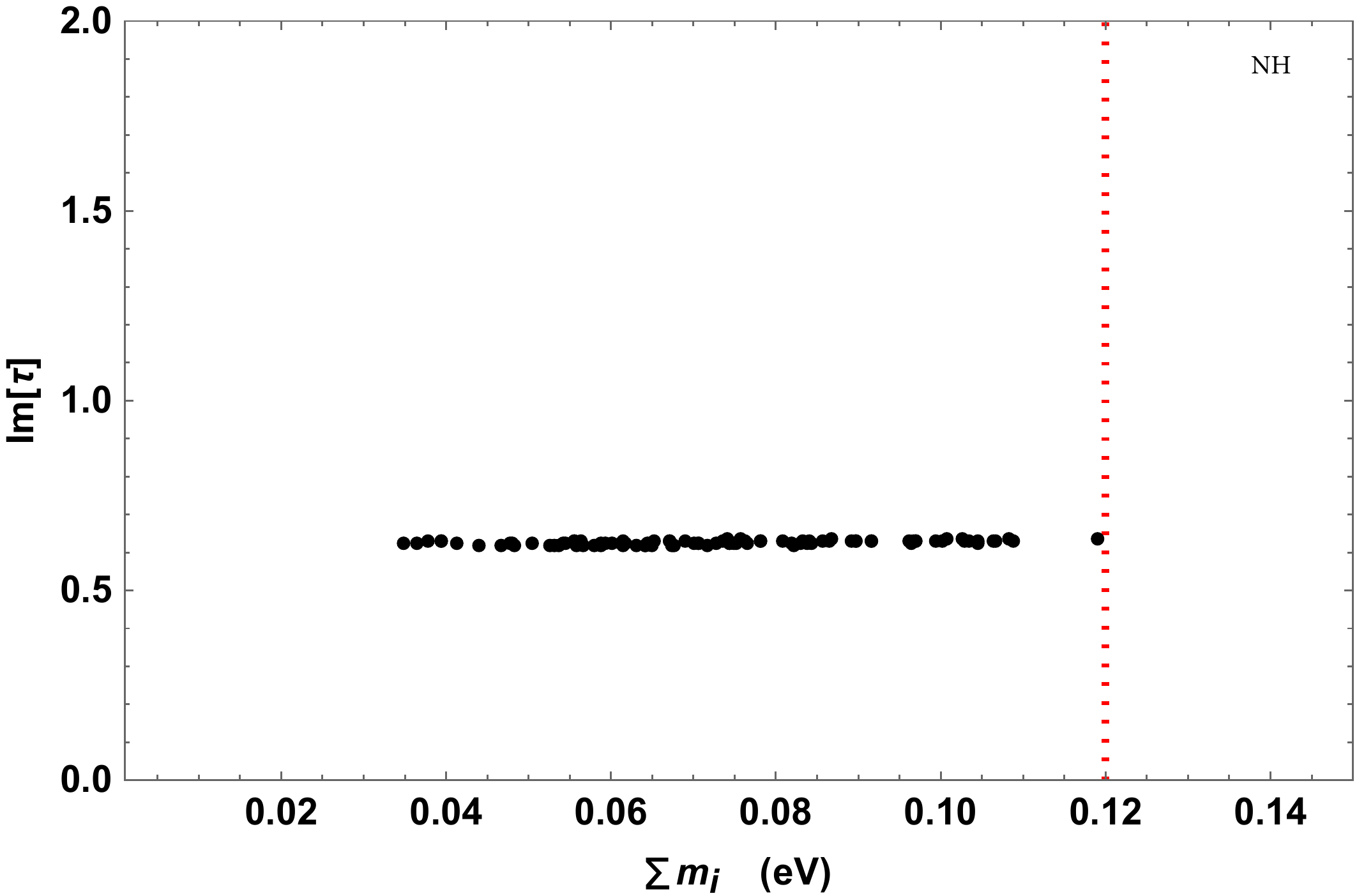}}
    \label{sumvsimtau}
    \quad
  %
  \caption{\footnotesize  The dependence of $\vert g\vert$  and Im$[\tau]$ on the sum of the active neutrino masses $\sum m_i$ for normal hierarchy.  }
  \label{C5gtauvssum}
\end{figure}

\begin{figure}
\centering
\subfigure[]{
    \includegraphics[width=0.45\textwidth]{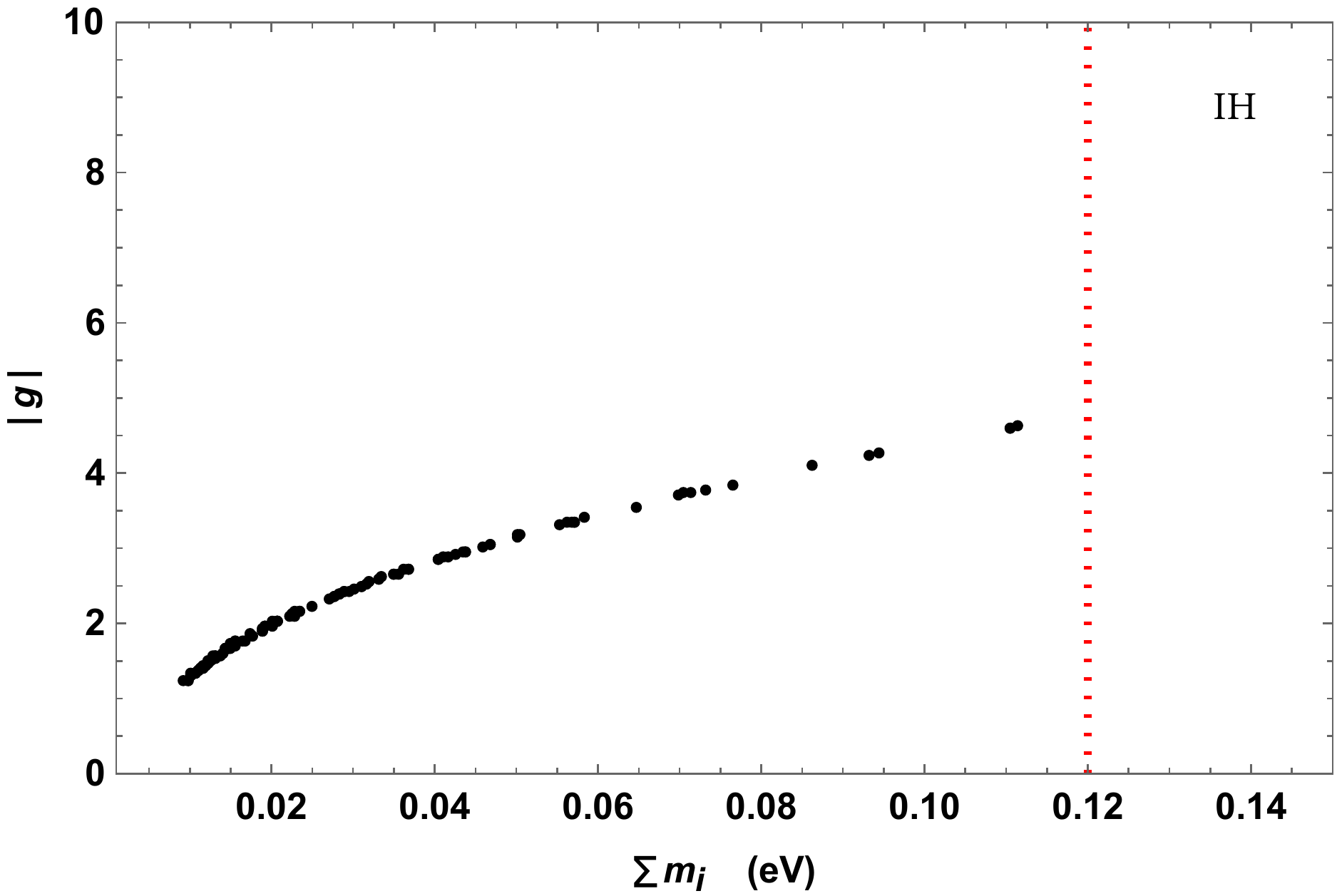}}
    \label{gvssum}
  \quad
\subfigure[]{
    \includegraphics[width=0.45\textwidth]{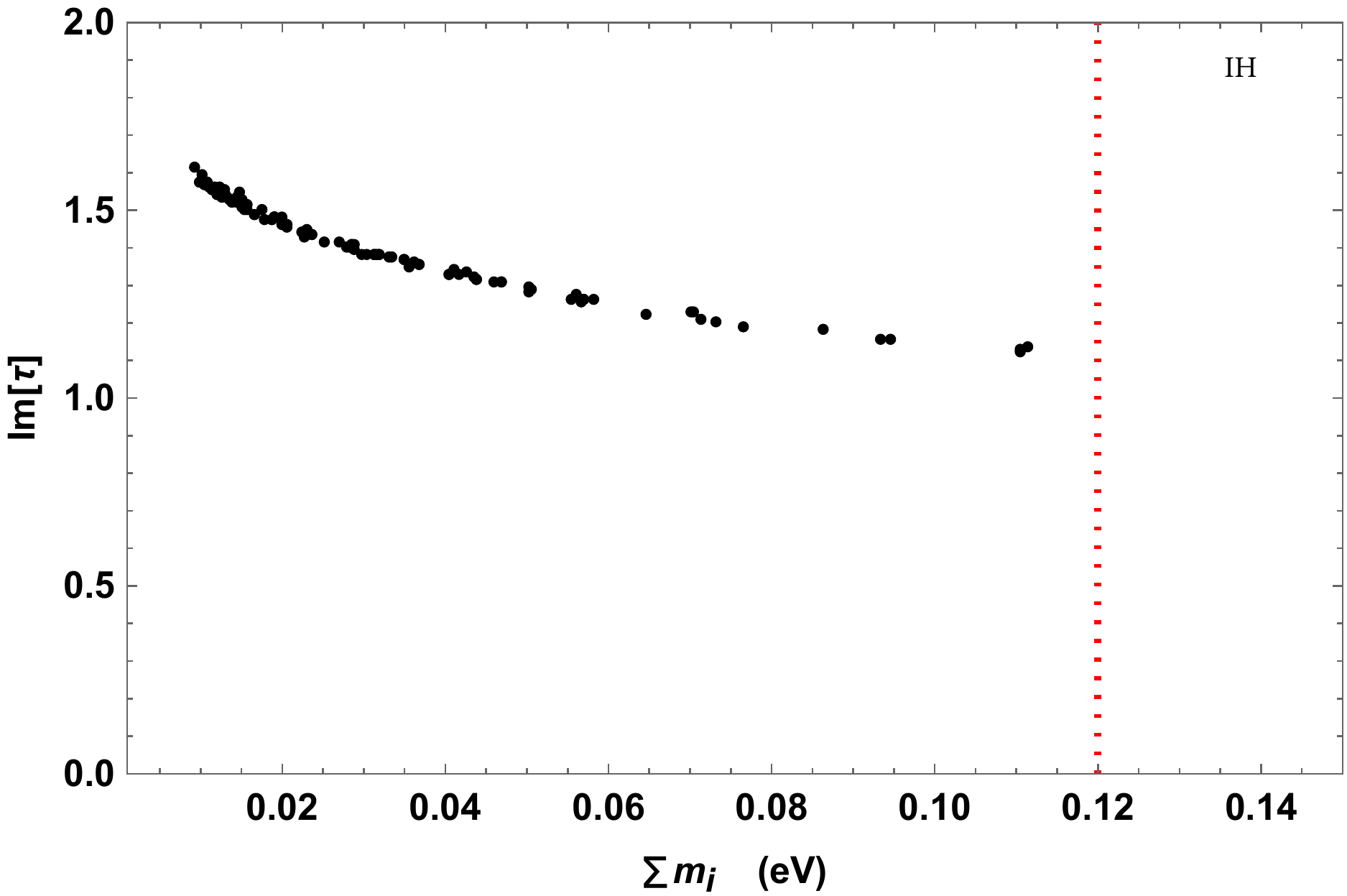}}
    \label{sumvsimtau}
    \quad
  %
  \caption{\footnotesize  The dependence of $\vert g\vert$  and Im$[\tau]$ on the sum of the active neutrino masses $\sum m_i$ for inverted hierarchy.  }
  \label{C5gtauvssumIH}
\end{figure}

For both the NH and IH, the allowed ranges of Re$[\tau]$ and Im$[\tau]$ are respectively shown in figure \ref{C5tauchargeplot}(b) and figure \ref{C5tauchargeplotIH}(b). The values of Im$(\tau)$ vary in the range 0.61 to 0.63 while the values of Re$(\tau)$ are continuously varied in a very small region between 0.32 to 0.41 for NH in the fundamental domain of $\tau$. Similarly, for IH, Im$(\tau)$ goes from 1.12 to 1.61 while Re$(\tau)$ varies in the range 0.01 to 0.14. The predicted ranges of Yukawa couplings $\vert y_1\vert,\ \vert y_2\vert$ and $\vert y_3\vert$, determined from these values, are shown in figure  \ref{yukawaplot} as variation plots for both the NH and IH. It is observed that the couplings lie in the ranges $0.79 \leq \vert y_1(\tau)\vert\leq 0.92 $, $ 1.44 \leq \vert y_2(\tau)\vert\leq 1.52 $ , and  $ 1.22 \leq \vert y_3(\tau)\vert\leq 1.32 $ for NH. On the other hand, the couplings lie in the ranges $0.993 \leq \vert y_1(\tau)\vert\leq 1.01 $, $ 0.19 \leq \vert y_2(\tau)\vert\leq 0.57 $ , and  $ 0.023 \leq \vert y_3(\tau)\vert\leq 0.16 $ for IH.

\begin{figure}
\centering
 \subfigure[]{
    \includegraphics[width=0.45\textwidth]{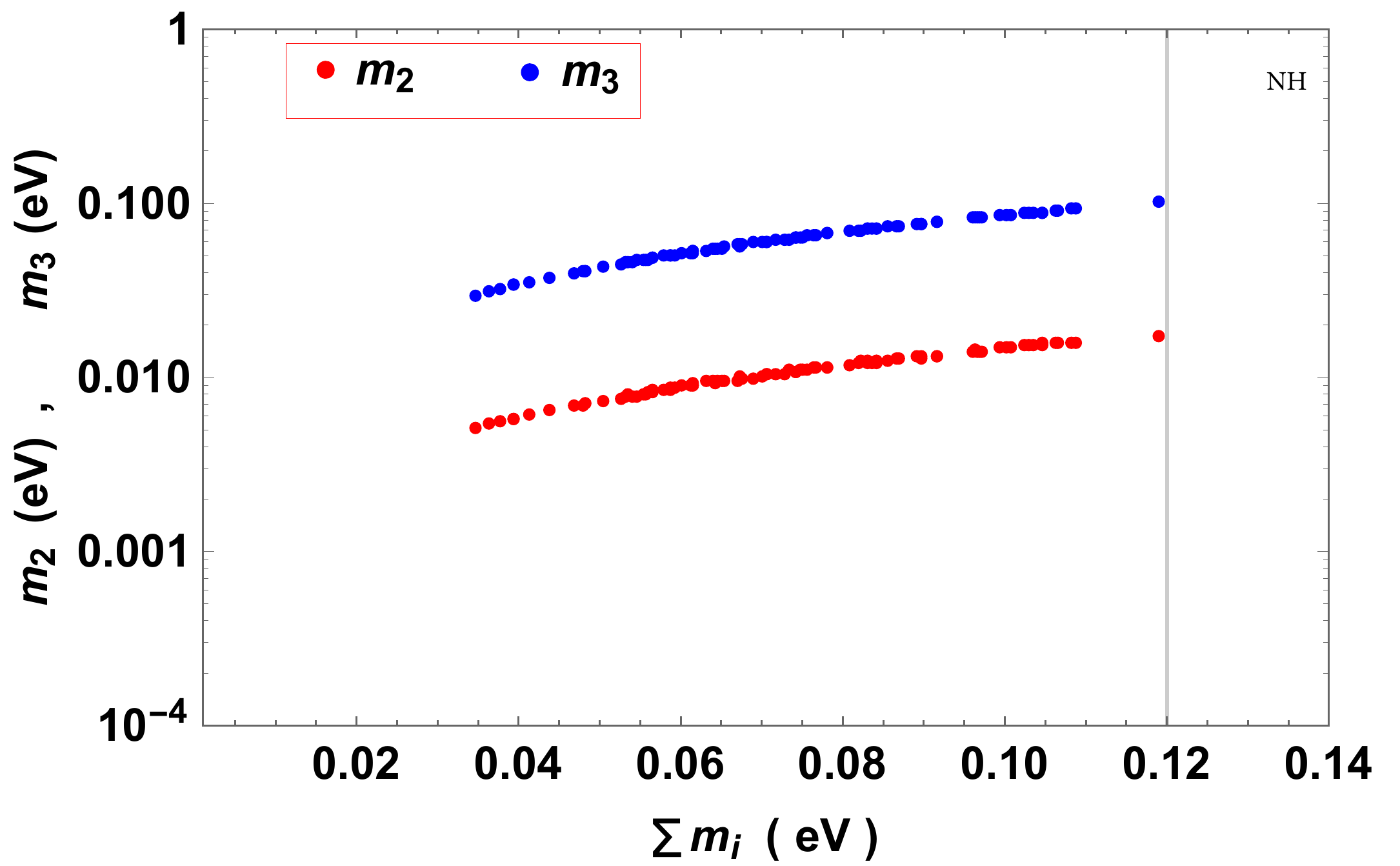}}
    \label{mass}
\quad
 \subfigure[]{
    \includegraphics[width=0.45\textwidth]{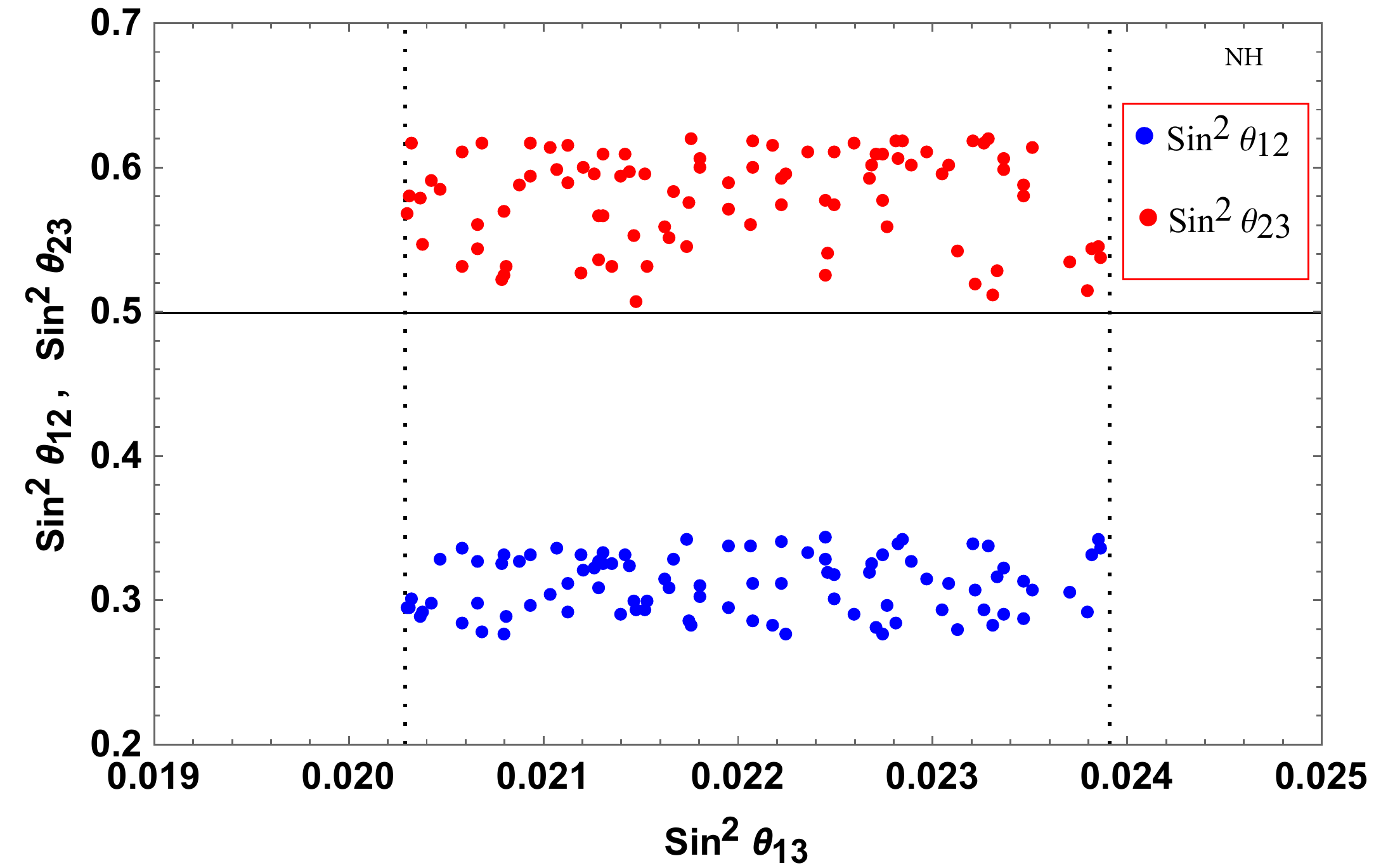}}
    \label{angles}
    \quad
  %
  \caption{\footnotesize (a) Predicted values of active neutrino mixing angles $\sin^2\theta_{12},\sin^2\theta_{23}$ and $\sin^2\theta_{13}$. (b) Predicted values of active neutrino masses $m_2,m_3$ ($m_1=0$) for normal hierarchy.  }
  \label{anglemass}
\end{figure}

\begin{figure}
\centering
 \subfigure[]{
    \includegraphics[width=0.47\textwidth]{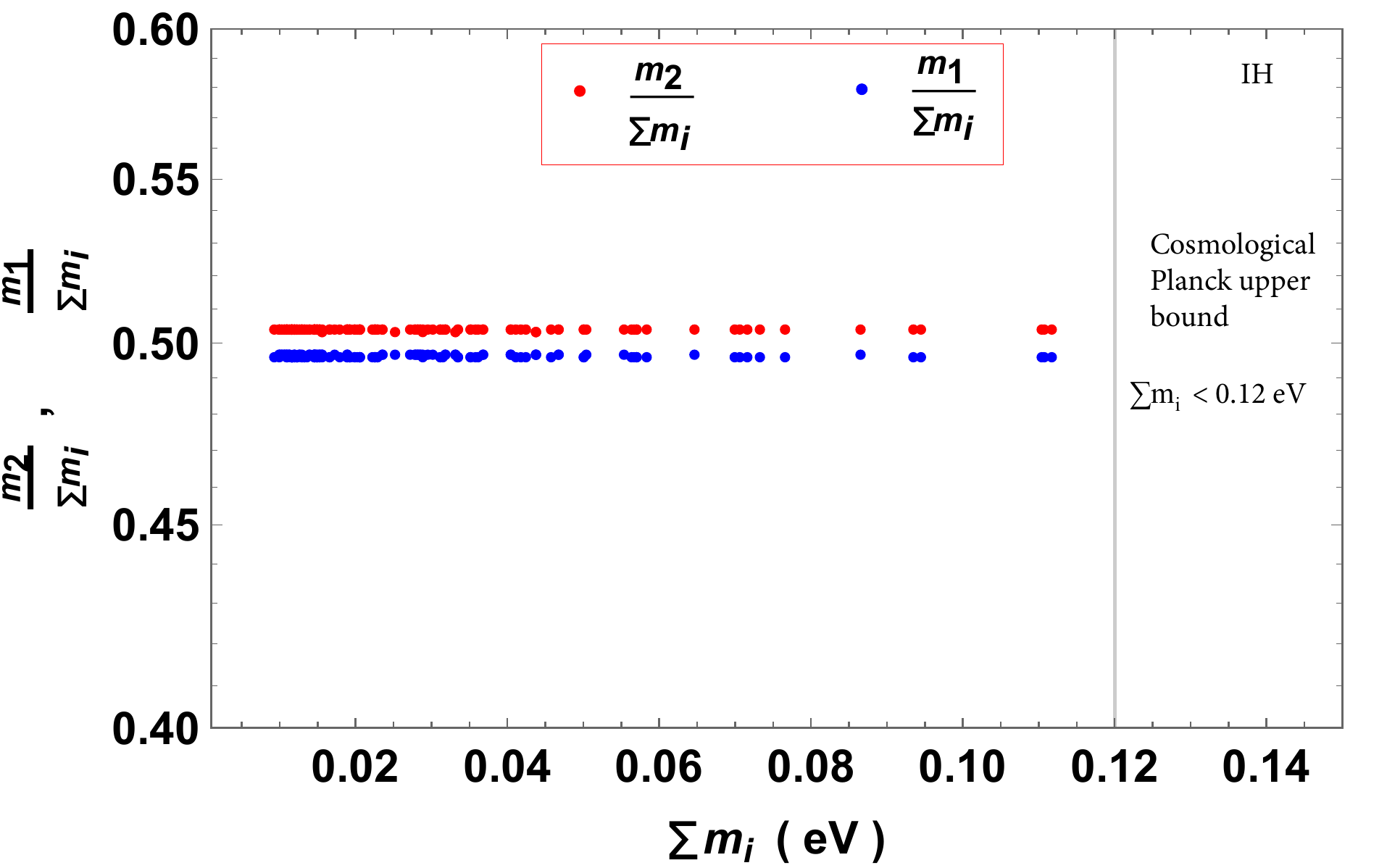}}
    \label{massIH}
\quad
 \subfigure[]{
    \includegraphics[width=0.45\textwidth]{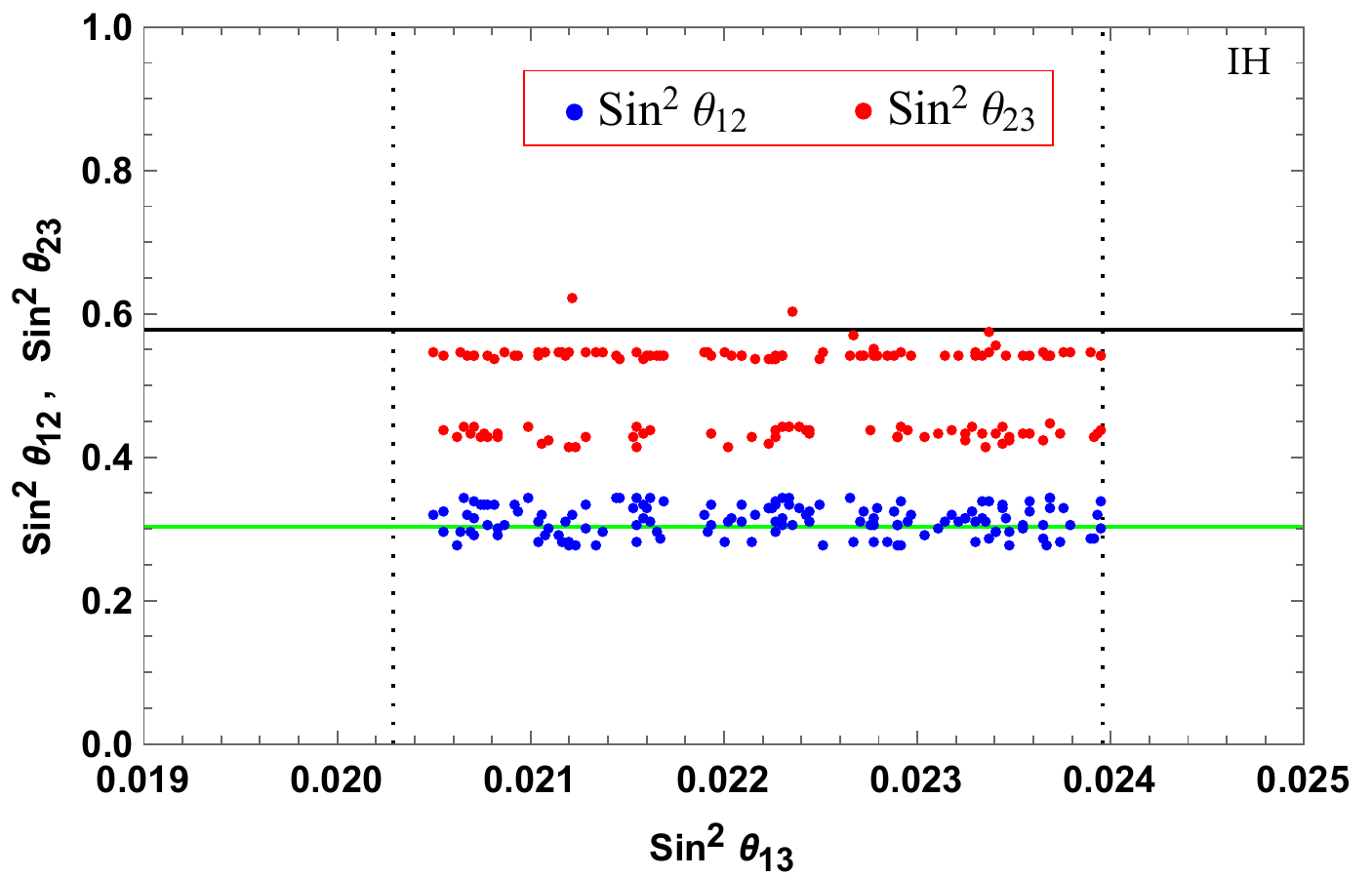}}
    \label{angles}
    \quad
  %
  \caption{\footnotesize (a) Predicted values of active neutrino mixing angles $\sin^2\theta_{12},\sin^2\theta_{23}$ and $\sin^2\theta_{13}$. (b) Predicted values of active neutrino masses $m_1,m_2$ ($m_3=0$) for inverted hierarchy.  }
  \label{anglemassIH}
\end{figure}

\section{Results of the analysis}\label{section4}
In this section, we shall show the results of the numerical analysis of the model. It is exciting to see in figure \ref{C5gtauvssum} and figure  \ref{C5gtauvssumIH} , which show the dependence of Im$[\tau]$ and $\vert g\vert$ with the sum of active neutrino masses $\sum m_i$ for NH and IH respectively. These figures show that $\vert g\vert$ and Im[$\tau$] are heavily constrained by the Planck upper bound $\sum m_i < 0.12$ eV, indicated by the vertical dotted line. The cosmological Planck upper limit on $\sum m_i$  is satisfied if we fix the overall scale factor $v_u^2g_1^2/\lambda = 0.000629$ eV for NH and $v_u^2g_1^2/\lambda = 0.00183 $ eV for IH in our analysis. Figures \ref{anglemass} and \ref{anglemassIH} show the variation of active neutrino masses  and active neutrino mixing angles for NH and IH respectively. These figures suggest that the mixing angles $\sin^2\theta_{12}$ and $\sin^2\theta_{13}$ run over their experimental ranges while the values of $\sin^2\theta_{23}$ are concentrated more in regions above 0.5 for IH. However, it needs to be more conclusive in this case to identify the octant degeneracy of $\theta_ {23}$. Whereas, in the case of NH, all the predicted values of $\sin^2\theta_{23}$ lie above 0.5 which suggest that higher octant of $\theta_{23}$ is predicted.

\begin{figure}
\centering
\subfigure[]{
    \includegraphics[width=0.45\textwidth]{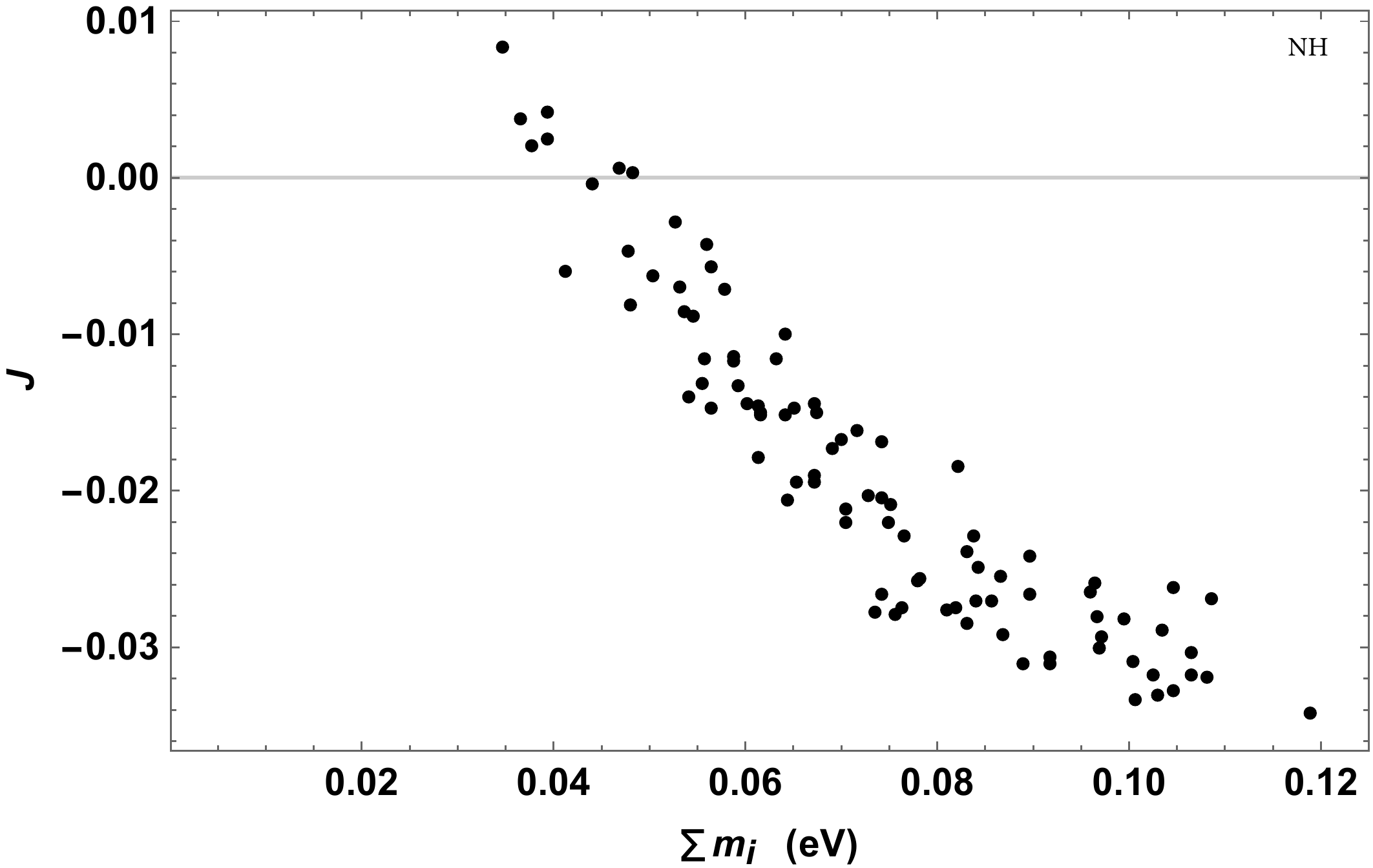}}
    \label{jplot}
\quad
\subfigure[]{
       \includegraphics[width=0.45\textwidth]{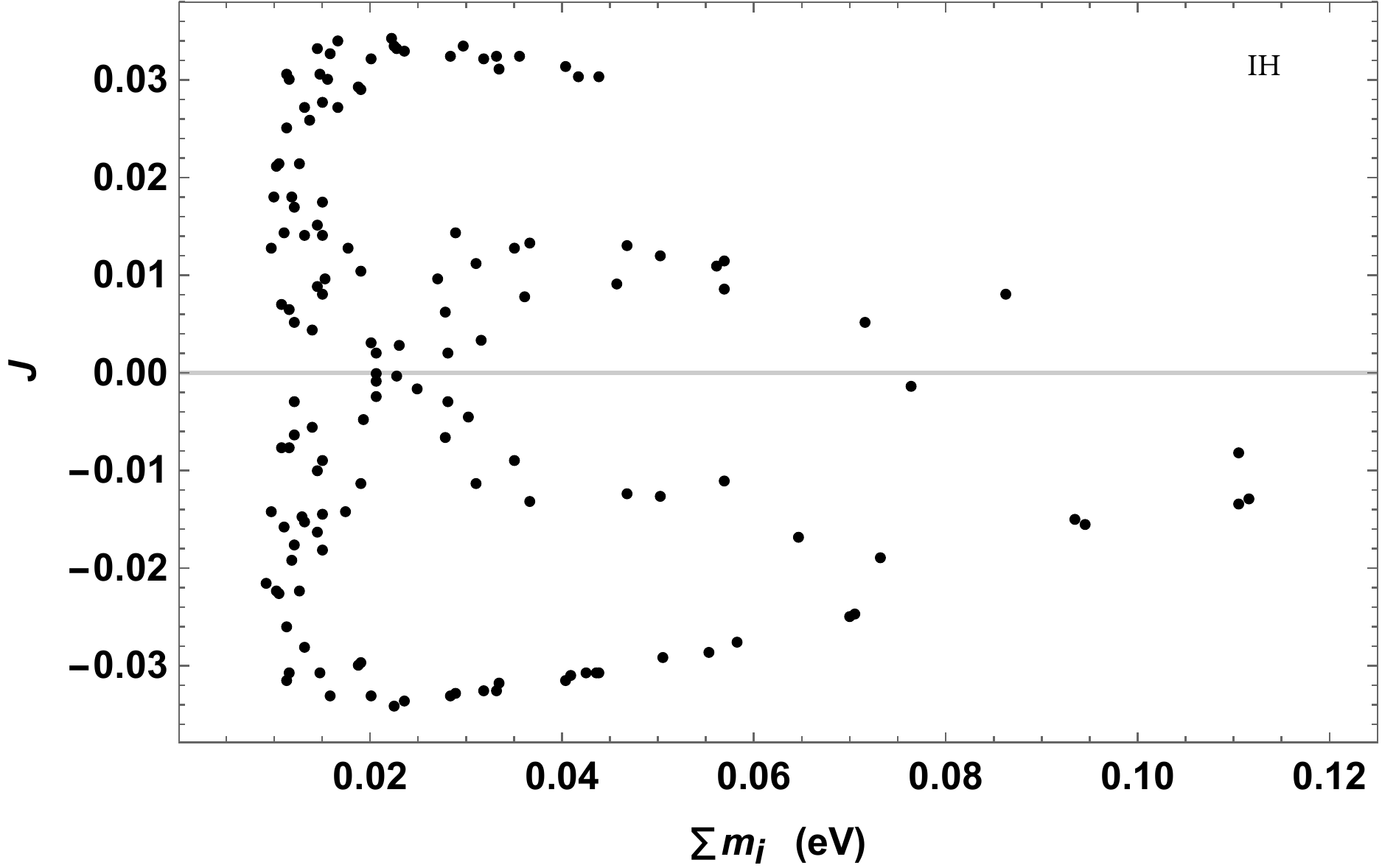}}
   \label{jplotih}
   \quad
  
  \caption{\footnotesize Figures show the model predictions of Jarlskog invariant $J$ normal hierarchy in (a) and inverted hierarchy in (b). }
  \label{C5jplot}
\end{figure}

\begin{figure}
\centering
\subfigure[]{
    \includegraphics[width=0.45\textwidth]{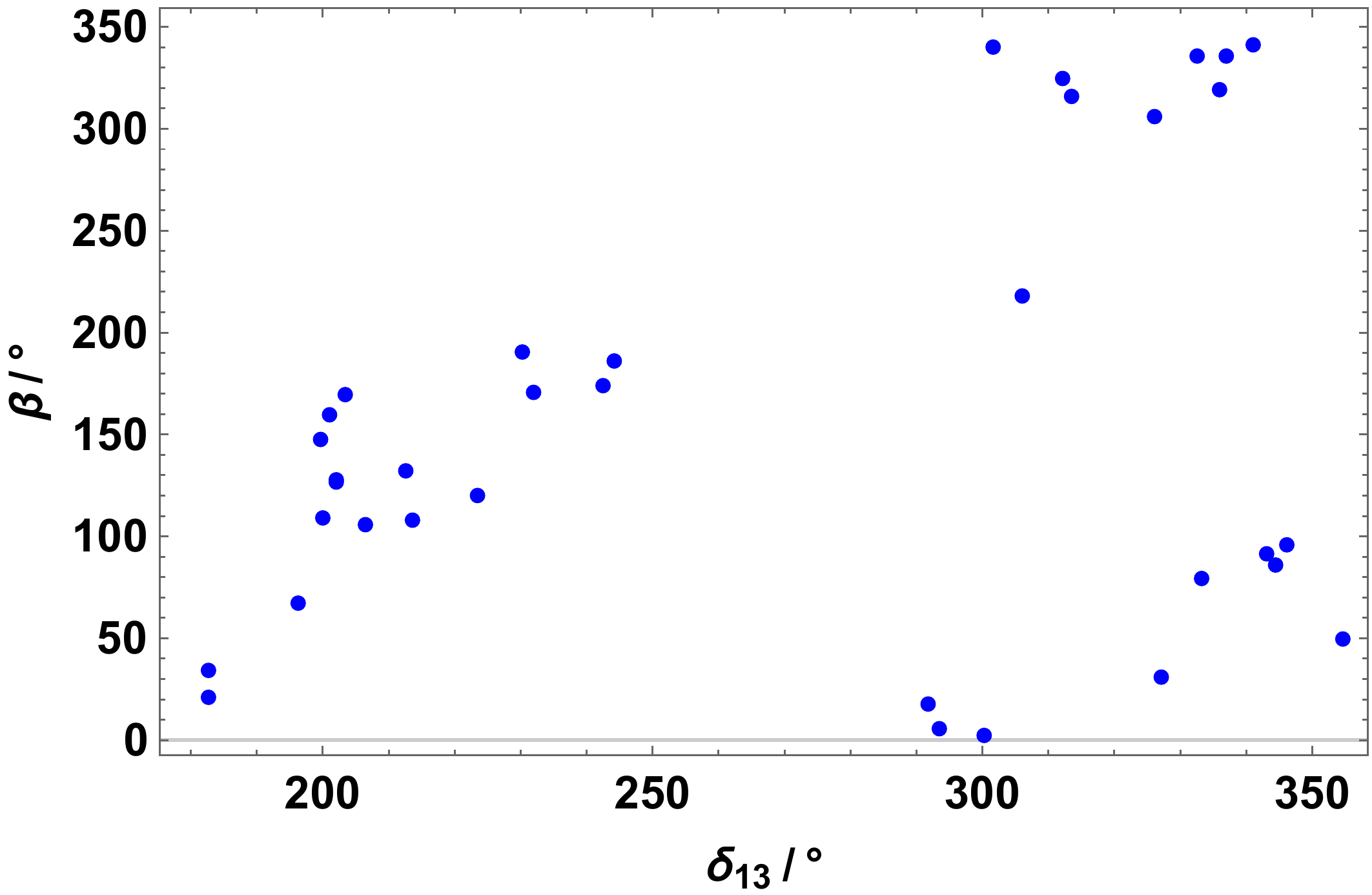}}
    \label{delphi2}
    \quad
\subfigure[]{
    \includegraphics[width=0.45\textwidth]{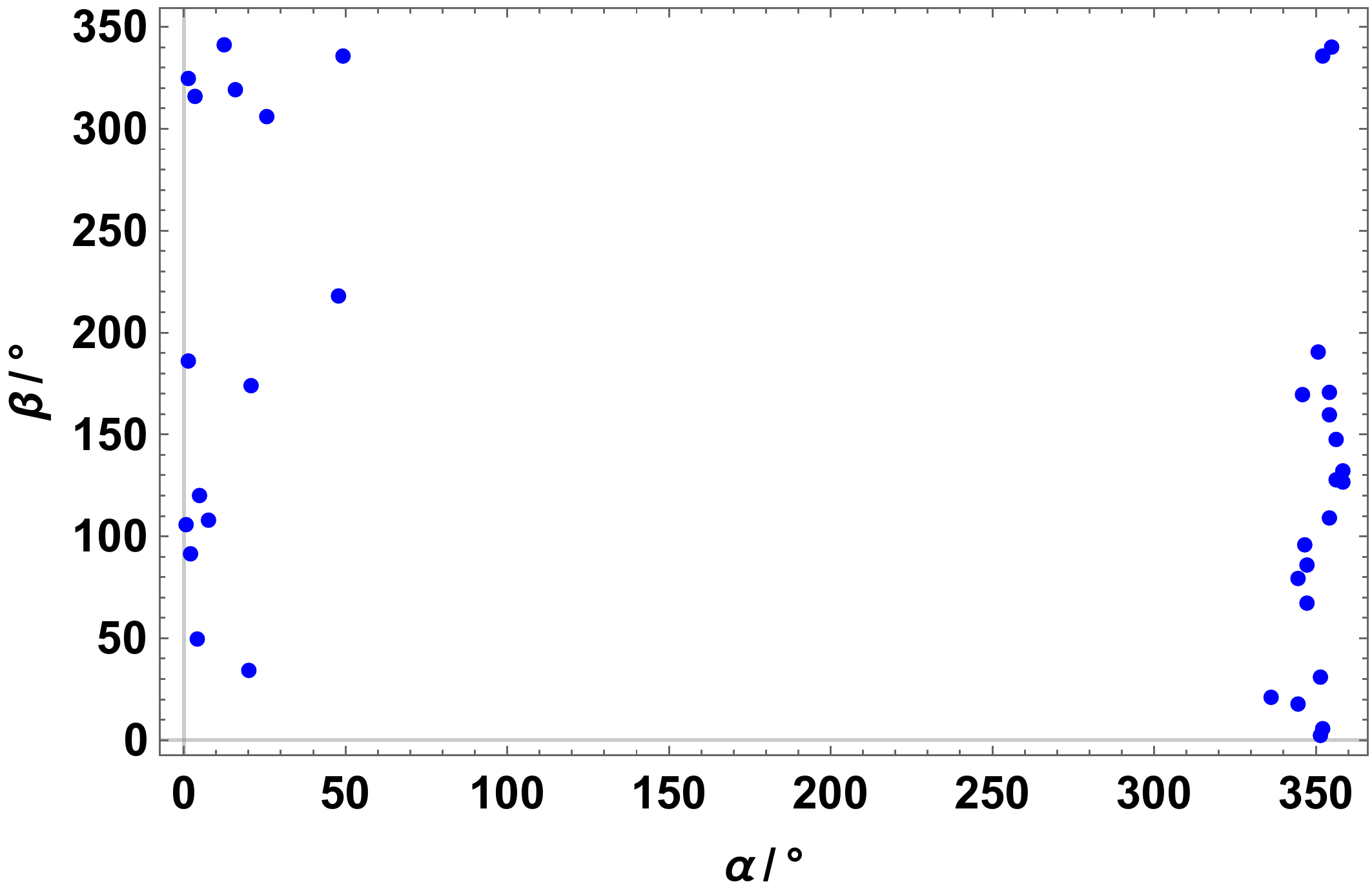}}
    \quad
   \label{phi1phi2}
  %
  \caption{\footnotesize Figures show the model predictions of CP-violating Dirac phase $\delta_{CP}$ and Majorana phases $\phi_1$, $\phi_2$ for normal hierarchy. }
  \label{C5phase}
\end{figure}

\begin{figure}
\centering
\subfigure[]{
    \includegraphics[width=0.45\textwidth]{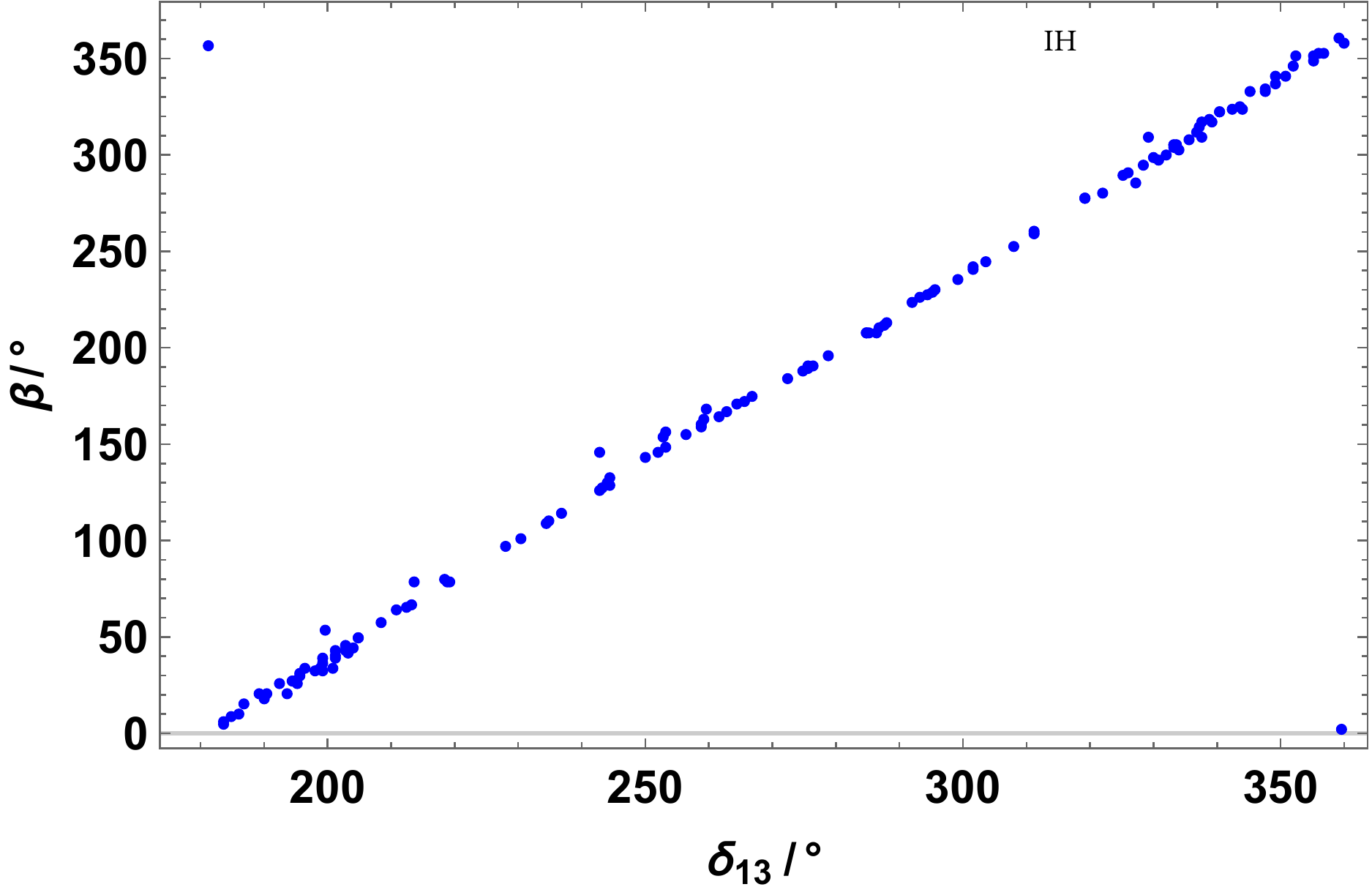}}
    \label{delphi2}
    \quad
\subfigure[]{
    \includegraphics[width=0.45\textwidth]{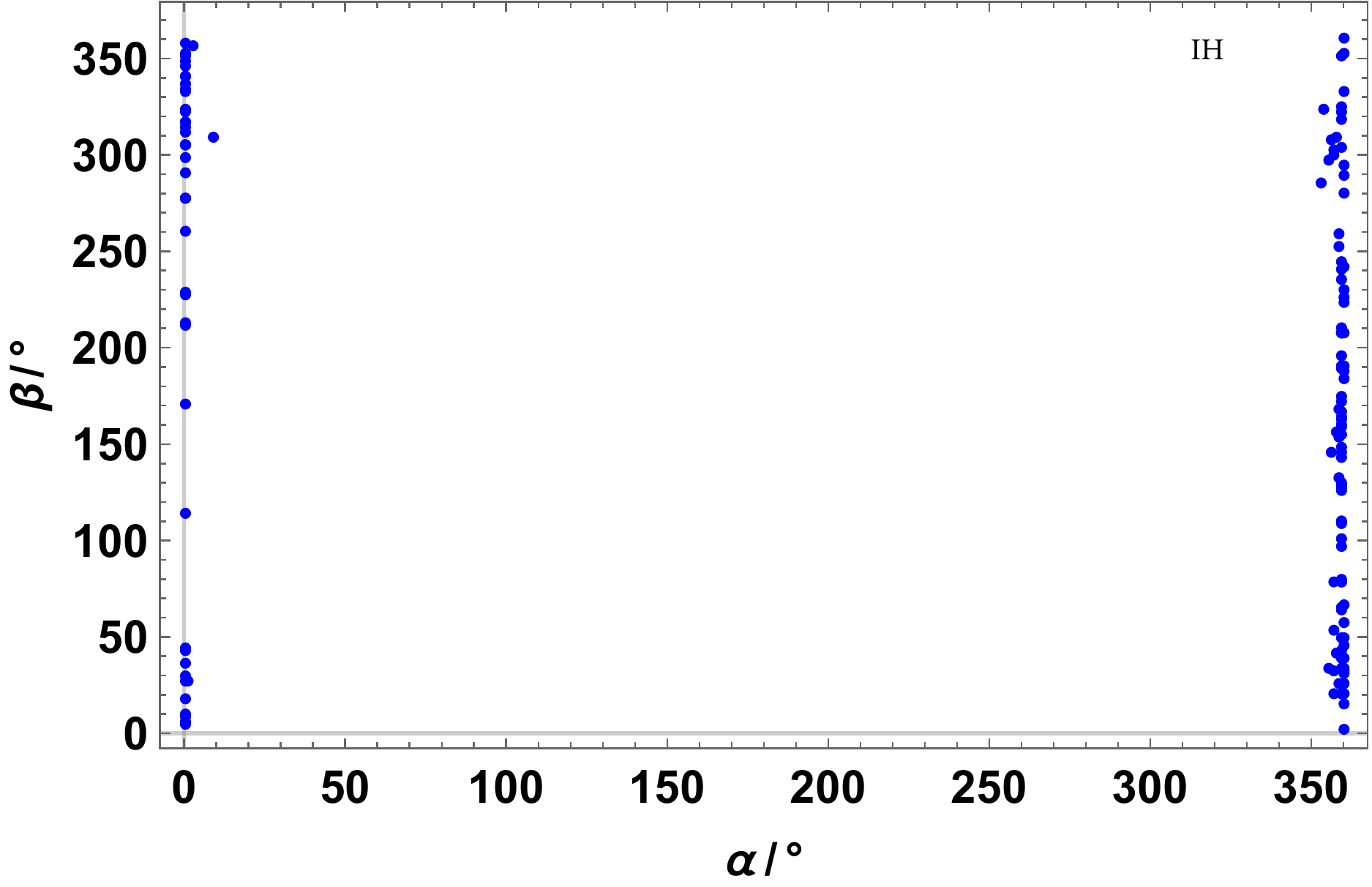}}
    \quad
   \label{phi1phi2}
  %
  \caption{\footnotesize Figures show the model predictions of CP-violating Dirac phase $\delta_{CP}$ and Majorana phases $\phi_1$, $\phi_2$ for inverted hierarchy. }
  \label{C5phaseIH}
\end{figure}

The variations of Jarlskog invariant $J$ with $\sum m_i$ are shown in figure \ref{C5jplot}. The model predictions of Dirac and Majorana CP-violating phases are solved from the invariants defined in (\ref{C5majoranaphase1}) - (\ref{C5majoranaphase2}), as shown in figure \ref{C5phase} - \ref{C5phaseIH}. In the MES mechanism, the lightest neutrino mass is always zero ($m_1 = 0$ for NH and $m_3=0$ for IH). It implies that one of the two Majorana phases is vanishing ($\alpha \simeq 0^o$ or $360^o$). From these figures, we can observe that the Majorana phase $\beta$ varies almost linearly with the Dirac CP-violating phase $\delta_{CP}.$ We also observe a vanishing $\alpha$ near $0^o$ or $360^o$ while $\beta$ varies continuously from $0^o $ to $ 360^o$ more profoundly in the case of IH in figure \ref{C5phaseIH}.

In the sterile neutrino sector, the input parameters given in (\ref{inputs}) evaluate the sterile neutrino in the eV mass scale. We evaluate the active sterile mixing elements. We also analyse the consistency of these results with the $\nu_{\mu} \rightarrow \nu_e$ appearance searches which are provided by the LSND and MiniBooNE. These results are extensively discussed in Ref. \citen{dasgupta2021sterile}. The effective mixing angle $\theta_{\mu e}$ for this case is defined as 
\begin{equation}
\sin^2 2\theta_{\mu e} \equiv 4 \vert U_{e4}\vert^2\vert U_{\mu 4}\vert^2.
\end{equation}
Similarly for the $\nu_e$ disappearance searches, the effective mixing angle is defined as 
\begin{equation}
\sin^2 2\theta_{ee} \equiv 4 \vert U_{e4}\vert^2(1- \vert U_{e 4}\vert^2).
\end{equation}
Finally, for the $\nu_{\mu}$ disappearance searches, the effective active sterile mixing angle is defined via 
\begin{equation}
\sin^2 2\theta_{\mu\mu} \equiv 4 \vert U_{\mu 4}\vert^2(1- \vert U_{\mu 4}\vert^2).
\end{equation}

\begin{figure}
\centering
\subfigure[]{
    \includegraphics[width=0.45\textwidth]{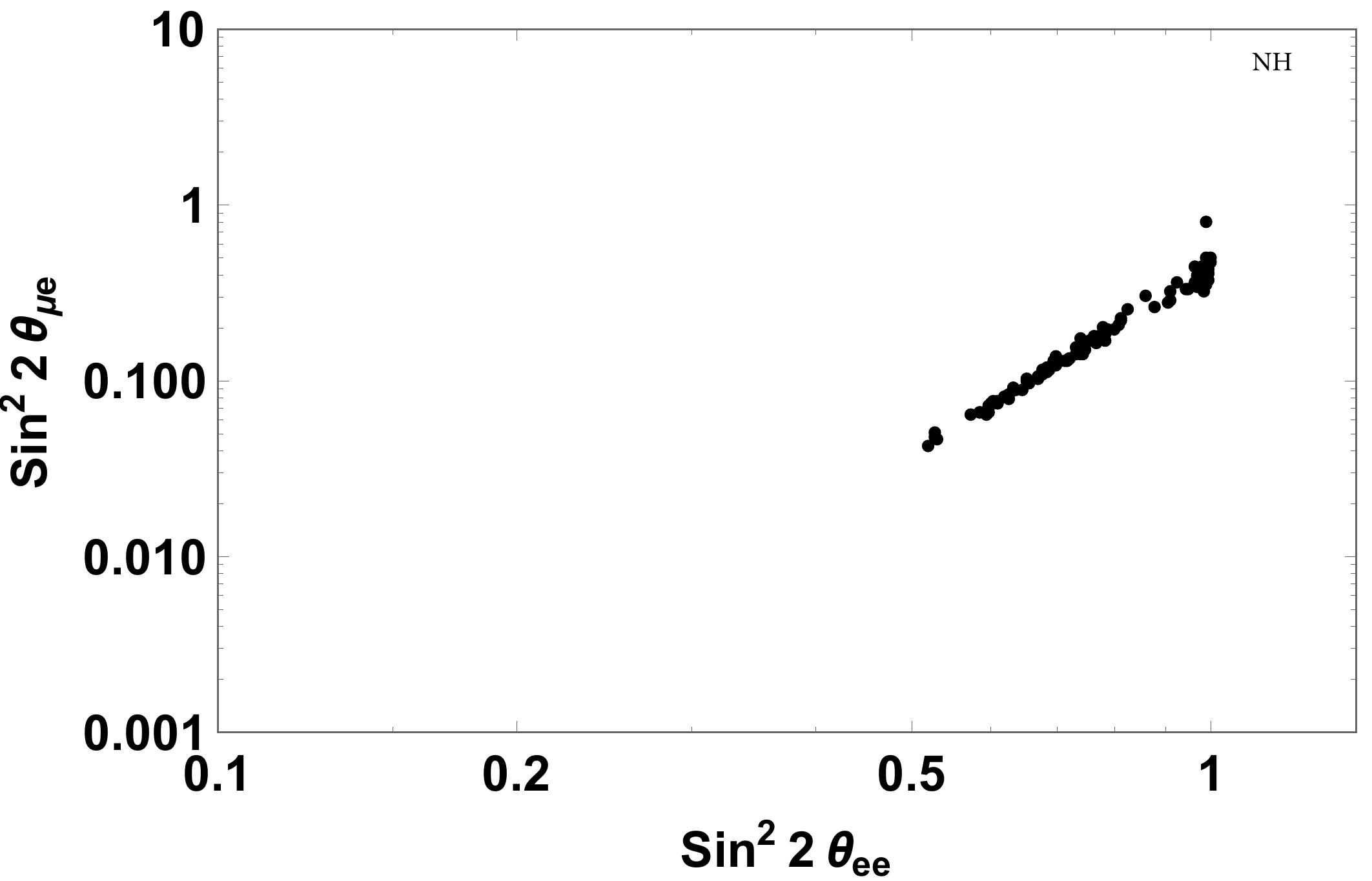}}
 \quad
 \subfigure[]{
 \quad
    \includegraphics[width=0.45\textwidth]{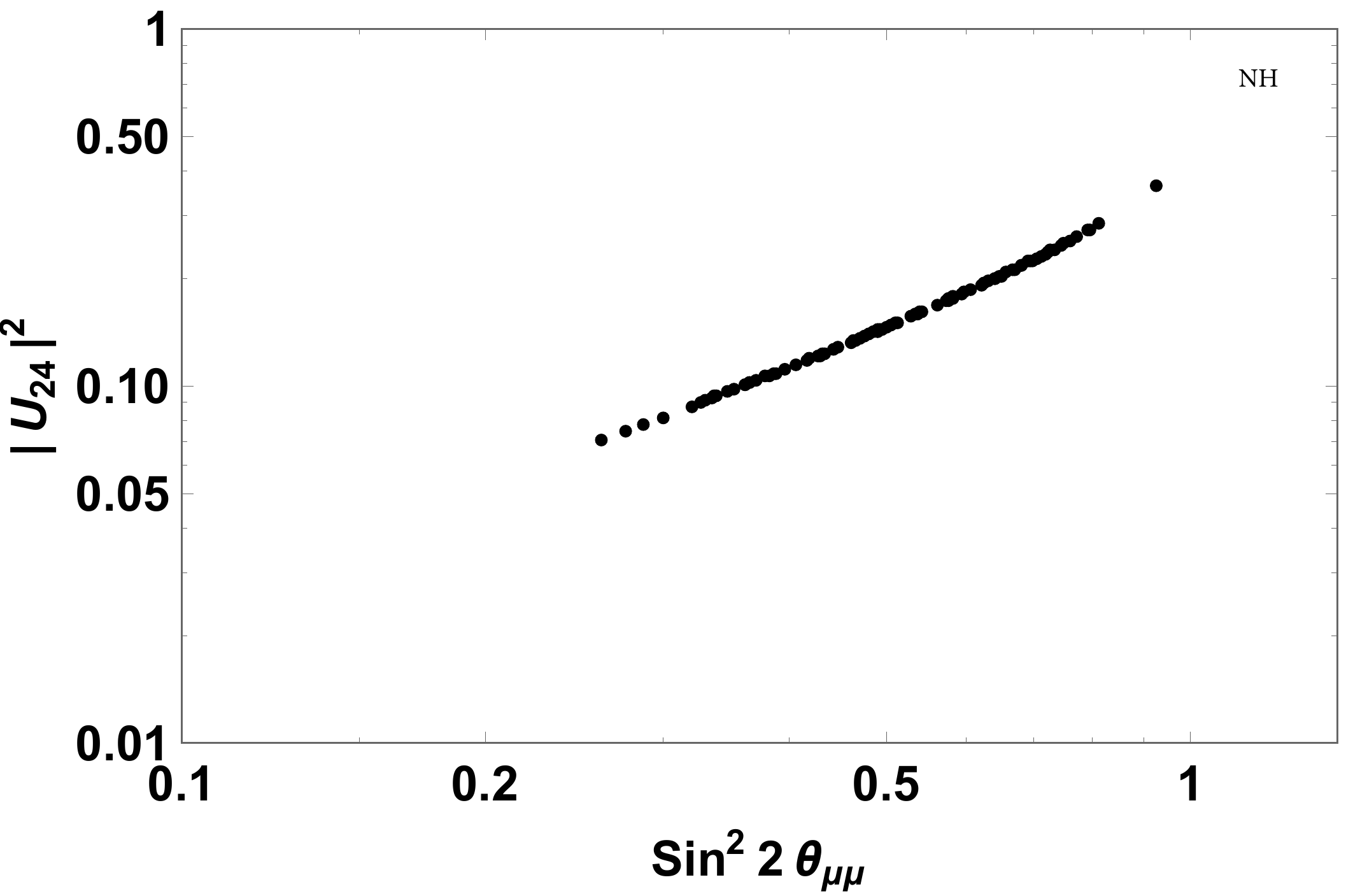}}
   
  %
  \caption{\footnotesize Figures show the variation of effective mixing angles $\sin^ 2\theta_{ee}$ with  $\sin^ 2\theta_{\mu\mu}$ in (a) and $\sin^ 2\theta_{\mu\mu}$ with $\vert U_{24}\vert^2$ in (b) for normal hierarchy.  }
  \label{effangleNH}
\end{figure}

\begin{figure}
\centering
\subfigure[]{
    \includegraphics[width=0.45\textwidth]{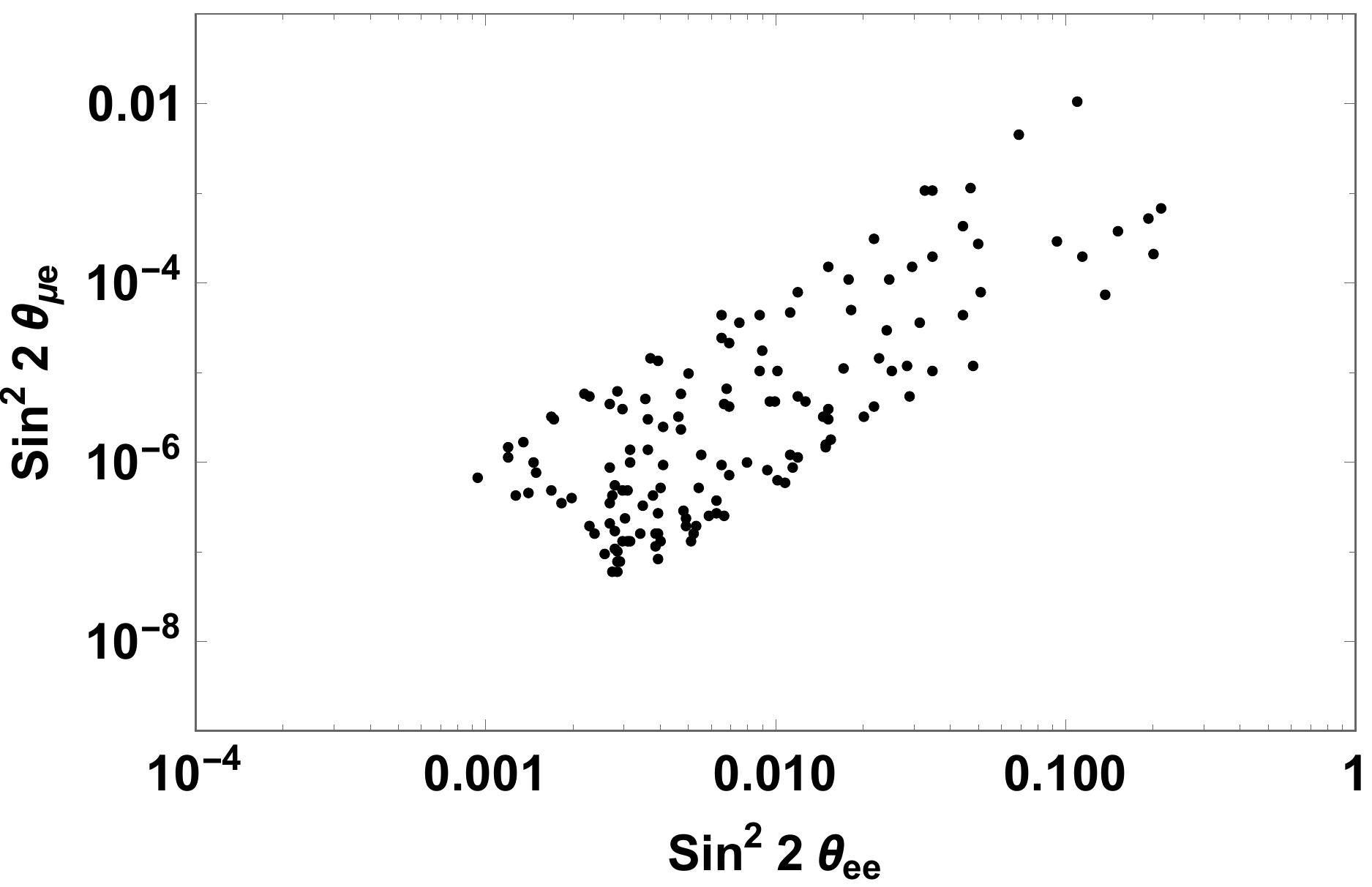}}
 \quad
 \subfigure[]{
 \quad
    \includegraphics[width=0.45\textwidth]{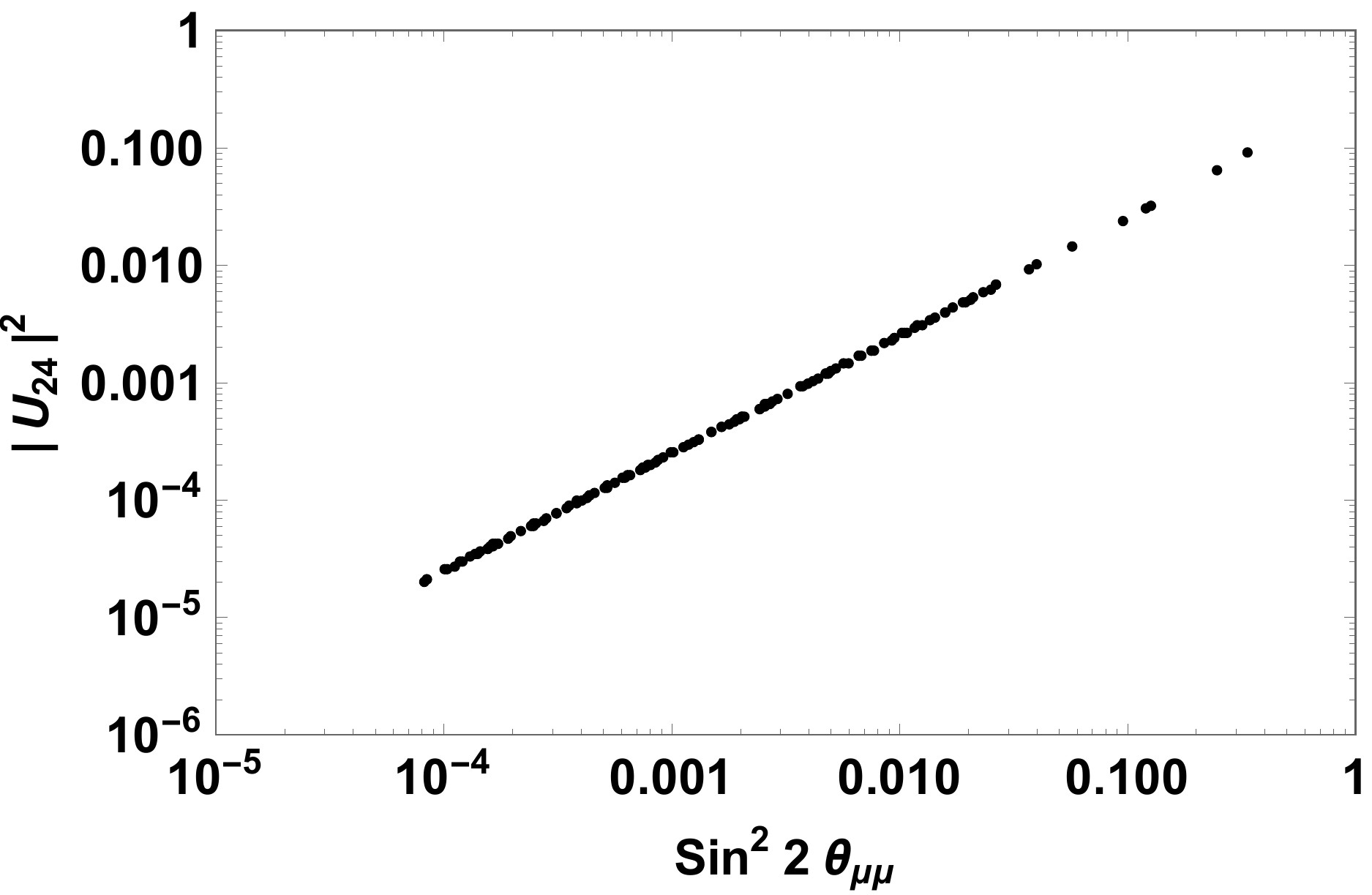}}
   
  %
  \caption{\footnotesize Figures show the variation of effective mixing angles $\sin^ 2\theta_{ee}$ with  $\sin^ 2\theta_{\mu\mu}$ in (a) and $\sin^ 2\theta_{\mu\mu}$ with $\vert U_{24}\vert^2$ in (b) for inverted hierarchy.}
  \label{effangleIH}
\end{figure}

The best fit region favored by MiniBooNE and LSND which is also in agreement with the null results from other experiments is around effective mixing angle $\sin^2 2\theta_{\mu e} \sim 0.01$ and the mass squared difference $\Delta m^2_{41} \lesssim 1$eV$^2$. In our analysis, we evaluate these effective mixing angles using the relations given above. The variation plots showing the effective mixing angles are shown in figure \ref{effangleNH} for NH and figure \ref{effangleIH} for IH respectively. These plots suggest that our analysis is consistent with the experimental results and global analysis \cite{dasgupta2021sterile}. The best-fit and 3$\sigma$ ranges of active-sterile mixing elements are also summarised in Table \ref{C5bestfit}.

\begin{figure}
\centering
\subfigure[]{
    \includegraphics[width=0.48\textwidth]{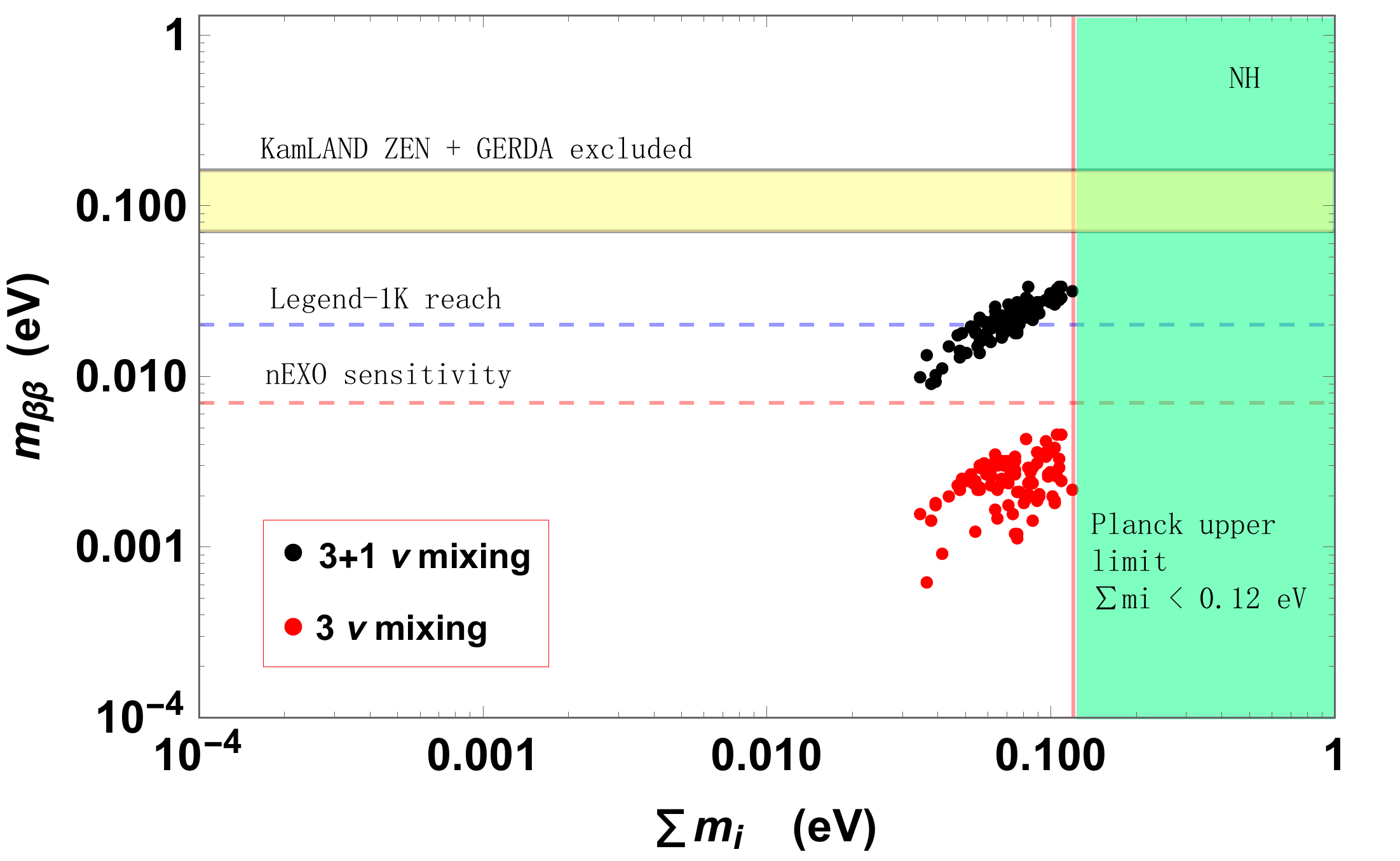}}
    \label{mbbplot}
\subfigure[]{
    \includegraphics[width=0.48\textwidth]{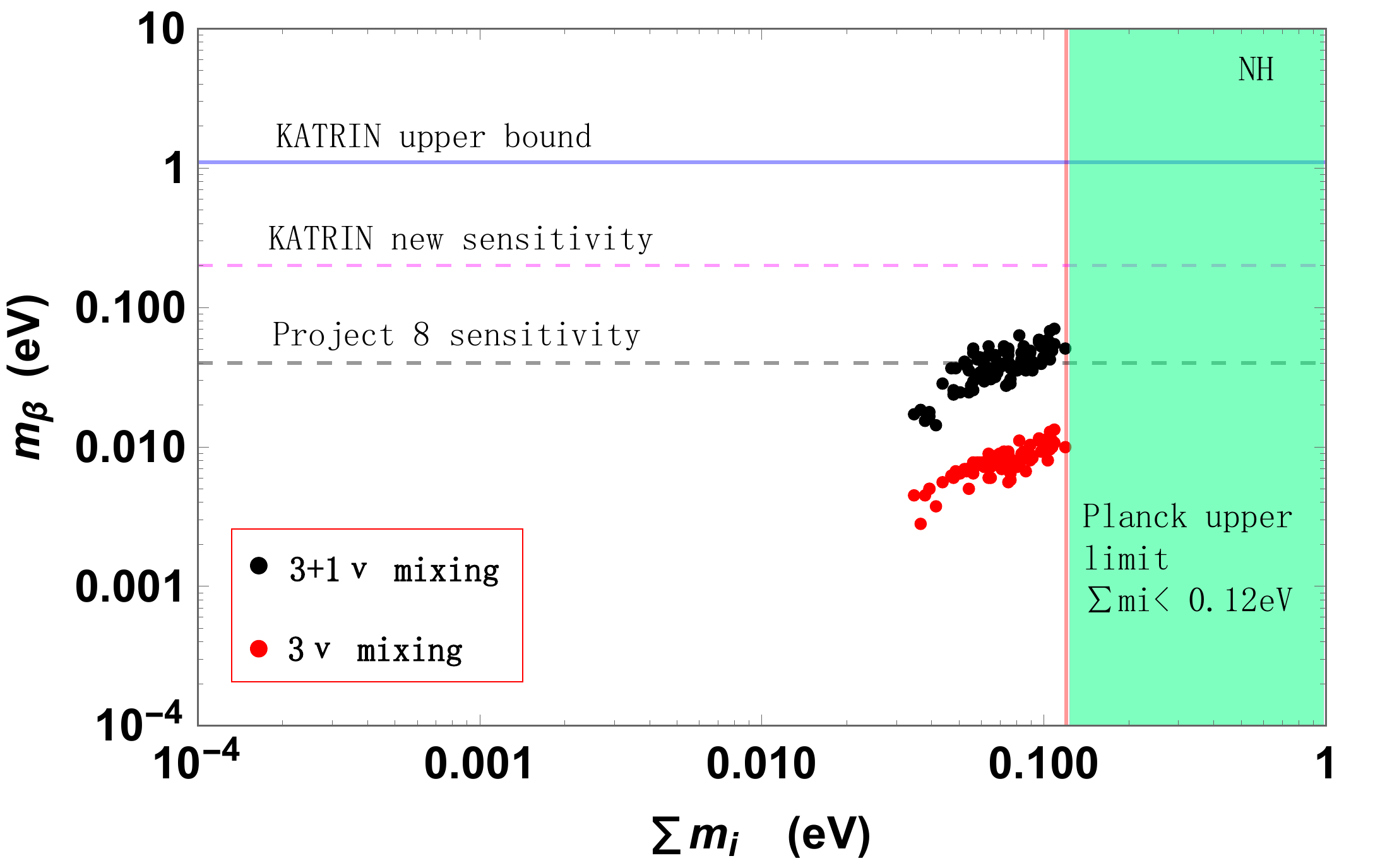}}
   \label{mbplot}
  %
  \caption{\footnotesize Dependence of effective mass parameters $m_{\beta\beta}$ and $m_{\beta}$ with $\sum m_i$ for NH. }
  \label{C5effmass}
\end{figure}

The model predictions of effective mass parameters $m_{\beta\beta}$ and $m_{\beta}$ are shown in figure \ref{C5effmass} for NH. The black data points represent the predictions for the 3+1 mixings, while the red data points are for the active neutrino sector. In the MES mechanism for NH, the contribution from $m_1$ vanishes as the lightest neutrino mass, $m_1=0$. There is a significant difference in the predictions for three neutrino and 3+1 neutrino mixings. It is observed that the presence of active-sterile mixings considerably increases the values of $m_{\beta}$ and $m_{\beta\beta}$. The dotted, horizontal red line in figure \ref{C5effmass}(a) indicates the future sensitivity of the nEXO experiment. Based on the present model, almost all the data points in 3+1 mixings are obtained inside nEXO and Legend-1K sensitivities, which have the potential to rule out or verify active-sterile mixings. However, predictions in three neutrino mixings still exceed this sensitivity, which could be explored by future experiments. Recently, data from KamLAND-Zen puts an upper bound on $m_{\beta\beta}$ in the range $m_{\beta\beta} \sim (28-122)$ meV. Predictions of $m_{\beta\beta}$ from our analysis which is observed in range $m_{\beta\beta} \sim (9.02 - 33.15)$ meV,  are still allowed by this upper bound.

In the case of $m_{\beta}$ in tritium beta decay, model predictions in 3+1 mixings are comparatively larger than active neutrino mixings. However, these results are still beyond the new KATRIN sensitivity of $m_{\beta} \sim 0.2$eV and the Project 8 sensitivity which is 0.04 eV.  The effective mass parameters are observed in the ranges $9.02\ \mbox{meV}\leq m_{\beta\beta}\leq 33.15\ \mbox{meV}$ and $14.37\ \mbox{meV} \leq m_{\beta} \leq 68.33 $ meV in the presence of active sterile neutrino mixing. 

\begin{figure}
\centering
\subfigure[]{
    \includegraphics[width=0.48\textwidth]{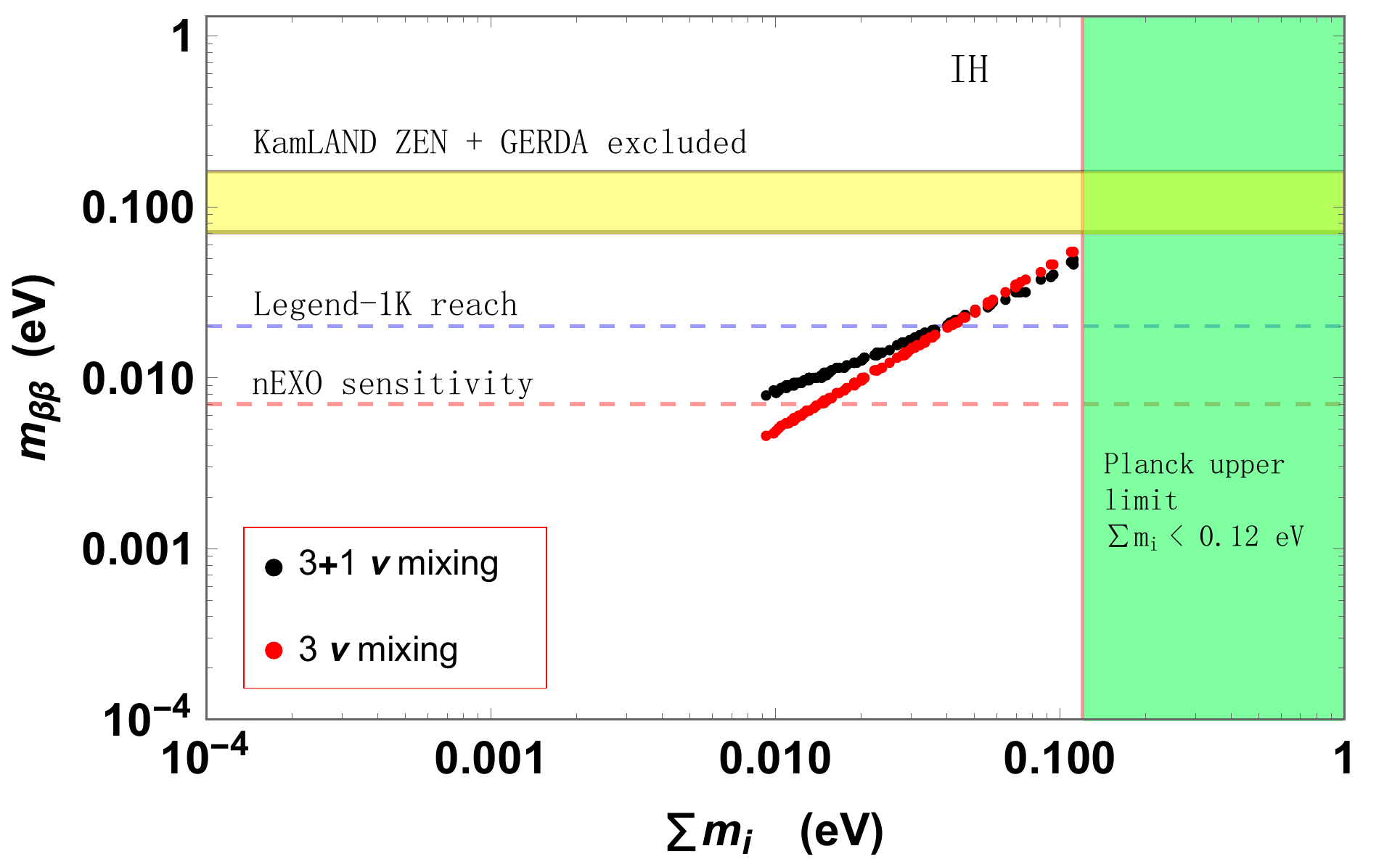}}
    \label{mbbplot}
\subfigure[]{
    \includegraphics[width=0.48\textwidth]{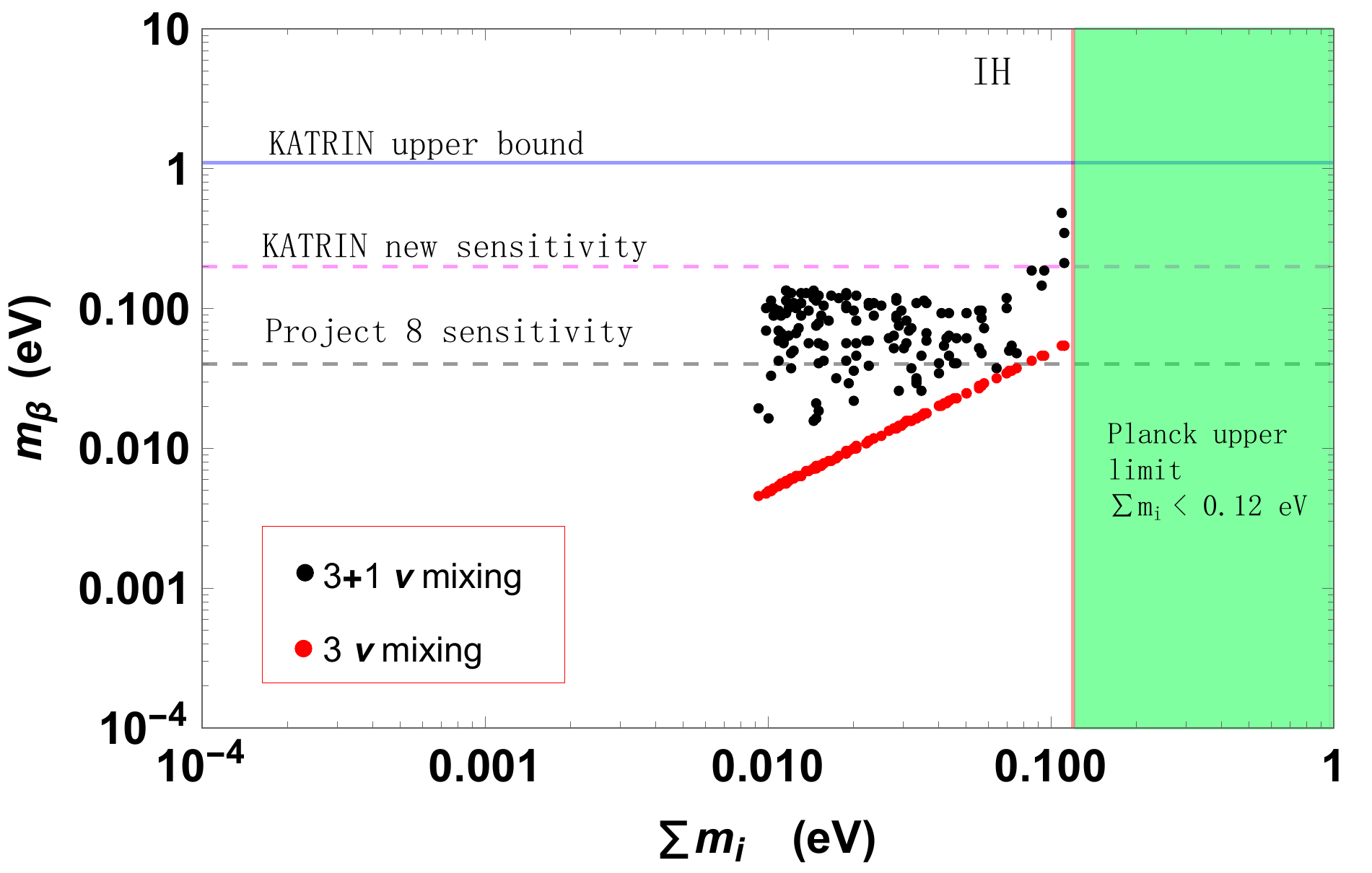}}
   \label{mbplot}
  %
  \caption{\footnotesize Dependence of effective mass parameters $m_{\beta\beta}$ and $m_{\beta}$ with $\sum m_i$ for IH. }
  \label{C5effmassIH}
\end{figure}

Similarly, for the case of IH, the effective mass parameters $m_{\beta\beta}$ and $m_{\beta}$ are shown in figure \ref{C5effmassIH}. In these plots, the predictions of $m_{\beta\beta}$ and $m_{\beta}$ are comparatively larger with respect to NH. However, all the data points are well below the excluded regions provided by experiments and our analysis is consistent with these data. The best fit results and 3$\sigma$ ranges of the effective mass parameters are summarised in Table \ref{C5bestfit}.

\begin{table}
\centering
\tbl{Best-fit values and 3$\sigma$ ranges of the model parameters and the corresponding predictions of neutrino observables for $\chi^2_{min} = 0.83 $ for NH and  $\chi^2_{min} = 3.61 $ for IH. Neutrino observables $\sin^2\theta_{12},\sin^2\theta_{13},\sin^2\theta_{23}$ and $r$ are constrained by 3$\sigma$ values from experimental data.}
{\begin{tabular}{@{}c|cccc @{}} 
\hline
\rule{0pt}{3ex} Neutrino observables & \multicolumn{2}{c|}{Best-fit} &  \multicolumn{2}{c} {3$\sigma$ range} \\

\cline{2-5}
& \rule{0pt}{3ex} NH & IH & NH & IH \\
\hline
\rule{0pt}{3ex}$\sin^2\theta_{23} $ & 0.573 & 0.550 & & - \\ 

$\sin^2\theta_{13}$ & 0.022 & 0.022 & - & -  \rule{0pt}{3ex}\\ 

$\sin^2\theta_{12}$ & 0.300 & 0.303 & - & - \rule{0pt}{3ex}\\ 

$r$ & 0.172 & 0.171 & - & -\rule{0pt}{3ex} \\ 
$\delta_{CP}/^{o}$ & 333.30 & 329.86 & [181.62 , 354.97] & [180.36 , 359.73]\rule{0pt}{3ex} \\
$\alpha /^{o}$ & 344.73 & 356.45 & [0 , 360] & [0 , 360] \rule{0pt}{3ex}\\
$\beta /^{o}$ & 79.56 & 295.83 & [0 , 360] & [0 , 360]\rule{0pt}{3ex} \\
$m_1/$ meV & 0  & 17.37 & 0 & [4.61 , 55.31]  \rule{0pt}{3ex}\\
$m_2/$meV & 9.05 & 17.63 & [5.16 , 17.37] & [4.68 , 56.17]   \rule{0pt}{3ex}\\
$m_3/$meV & 52.55 & 0 & [29.62 , 101.52] & 0   \rule{0pt}{3ex}\\
$m_4/$ meV & 82.28 & 181.72 & [19.78 , 158.53] & [59.43 , 11793.60]  \rule{0pt}{3ex}\\
$\sum$ $m_i $/meV & 61.60  & 35.00 & [34.78 , 118.90] & [9.29, 111.49] \rule{0pt}{3ex} \\
$m_{\beta}/$ eV & 0.037 & 0.256 & [0.014 , 0.068] & [0.01 , 0.48] \rule{0pt}{3ex}\\
$m_{\beta\beta}/$eV & 0.017 & 0.018 & [0.009 , 0.033] & [0.007 , 0.049] \rule{0pt}{3ex} \\
$\vert U_{14}\vert^2$ & 0.205 & 0.011 & [0.152 , 0.548] & [0.23 , 56.69 ]$\times 10^{-3}$ \rule{0pt}{3ex}\\
$\vert U_{24}\vert^2$ & 0.123 & 0.009 & [0.070 , 0.363] & [0.02 , 92.02]$\times 10^{-3}$ \rule{0pt}{3ex}\\
$\vert U_{34}\vert^2$ & 0.106 & 0.085 & [0.078 , 0.319] & [0.48 , 334.17]$\times 10^{-3}$ \rule{0pt}{3ex}\\
\hline 
\rule{0pt}{3ex}
Model parameters & \multicolumn{2}{c|}{Best-fit} & \multicolumn{2}{c}{3$\sigma$ range} \\
\cline{2-5}
& \rule{0pt}{3ex} NH & IH & NH & IH \\
\hline
\rule{0pt}{3ex} Re$[\tau]$ & 0.362 & 0.032 & [0.32 , 0.41] & [0.01 , 0.14]  \\ 
Im$[\tau]$ & 0.62 & 1.36 & [0.61 , 0.63] & [1.12 , 1.61] \rule{0pt}{3ex} \\ 
$\alpha^{\prime}$ & 2.82 $\times 10^{-4}$  & 5.72$\times 10^{-4}$  & [2.80 , 2.90]$\times 10^{-4}$ & [5.17 , 99.87] $\times 10^{-4}$  \rule{0pt}{3ex} \\  
$\beta^{\prime}$ & 1.69$\times 10^{-6}$ & 9.62$\times 10^{-3}$ & [1.57 , 1.96]$\times 10^{-6}$ & [5.54 , 100.03]$\times 10^{-4}$ \rule{0pt}{3ex} \\ 
$\gamma^{\prime}$ & 4.74 $\times 10^{-3}$& 3.64$\times 10^{-6}$ & [4.71 , 4.87]$\times 10^{-3}$ &  [3.10 , 7.22] $\times 10^{-6}$ \rule{0pt}{3ex} \\
$\vert g\vert$ & 3.72 & 2.65 & [2.54 , 5.45] & [1.24 , 4.62] \rule{0pt}{3ex}\\ 
\hline 
\end{tabular} 
\label{C5bestfit}}
\end{table}

\subsection{$\chi^2$ analysis}\label{C}
Finally, the best-fit values of the neutrino observables and the corresponding best-fit values of model parameters $\tau$ and $g$ are evaluated using the $\chi^2$ analysis. We use the $\chi^2$ function defined as 
\begin{equation}
\chi^2(x_i) = \sum_{j}\left(\frac{y_j(x_i)-y_j^{bf}}{\sigma_j}\right)^2,
\label{chitest}
\end{equation}
where $x_i$ are the free parameters in the model and $j$ is summed over the observables $\{\sin^2\theta_{12},\ \sin^2\theta_{13},\ \sin^2\theta_{23},r\}$. Here, $y_j(x_i)$ denotes the model predictions for the observables, and $y_j^{bf}$ are their best-fit values obtained from the global analysis. $\sigma_j$ denotes the corresponding uncertainties obtained by symmetrizing $1\sigma$ range of the neutrino observables given in Table \ref{data}. By minimizing the $\chi^2$ function, we can calculate the best-fit values of our model parameters and predict the values of neutrino observables.

For NH, at $\chi^2_{min}=0.83$, the best values of the free parameters of the model are found at Re$[\tau]= 0.362$, Im$[\tau]= 0.62$ and $\vert g\vert = 3.72$. The corresponding best-fit values of the neutrino observables are $\sin^2\theta_{23}=0.573,$ $\sin^2\theta_{12}=0.300,$ $ \sin^2\theta_{13}=0.022$ and $r = 0.172$. Lastly, Dirac and Majorana phases are observed at $\delta_{CP}=333.30^o,\ \alpha=344.73^o$ and $\beta=79.56.10^o$.  Whereas, for IH, at $\chi^2_{min}=3.61$, the best values of the free parameters are obtained at Re$[\tau]= 0.032$, Im$[\tau]= 1.36$ and $\vert g\vert = 2.65$. The corresponding best-fit values of the neutrino observables are $\sin^2\theta_{23}=0.550,$ $\sin^2\theta_{12}=0.303,$ $ \sin^2\theta_{13}=0.022$ and $r = 0.171$. Finally, the Dirac and Majorana phases are observed at $\delta_{CP}=329.86^o,\ \alpha=356.45^o$ and $\beta=295.83.10^o$. 

We have also evaluated the best-fit predictions of the $(3\times 3$) active neutrino mixing matrix for both NH and IH as shown below,

\begin{eqnarray}
\vert U\vert^{NH}_{bf} = \left(
\begin{array}{m{1.3cm} m{1.3cm} c}
 0.826 & 0.542 & 0.149 \\
 0.269 & 0.605 & 0.748 \\
 0.493 & 0.582 & 0.645  \\
\end{array}
\right),
\end{eqnarray}
and 
\begin{equation}
\vert U\vert^{IH}_{bf} = \left(
\begin{array}{m{1.3cm} m{1.3cm} c}
 0.827\hspace{2pt}  & 0.542 \hspace{2cm} & 0.147 \\
 0.459\hspace{2cm} & 0.509 \hspace{2cm} & 0.727 \\
 0.322 \hspace{2cm} & 0.668\hspace{2cm} & 0.670 \\
\end{array}
\right).
\end{equation}

The best-fit active-sterile mixing matrix $R$ for NH is given by 
\begin{eqnarray}
R_{bf} = \left(
\begin{array}{c}
 0.436694 - 0.120716 i \\
 0.349686 + 0.0356031 i \\
- 0.191268 - 0.26455 i \\
\end{array}
\right).
\end{eqnarray}

For IH, it is given by 
\begin{eqnarray}
R_{bf} = \left(
\begin{array}{c}
 0.0308322\, - 0.0233975 i \\
 - 0.0253386 - 0.0326925 i \\
- 0.0760387 - 0.0718129 i \\
\end{array}
\right).
\end{eqnarray}

We have summarized the best-fit values of the model parameters and neutrino observables in Table \ref{C5bestfit}. These results clearly suggest that the model predicts the higher octant of $\theta_{23}$ in NH. However, in the case of IH, it is not conclusive since some of the data points are obtained in the lower half of 0.5 with more data concentration in the upper half. Thus, the model still favours higher octant of $\theta_{23}$ in IH.

\section{Summary and Discussion}\label{section6}
We have successfully constructed a new neutrino mass model based on modular $A_4$ by extending the SM with an $A_4$ triplet right-handed neutrino $N$ and a singlet sterile neutrino $S$ in the 3+1 scheme. The model is found to give both the normal and inverted mass hierarchies at 3$\sigma$ for the parameter space given in (\ref{pspace}) in separately distinct regions of $\tau$. Our primary motivation is to avoid the hypothetical scalar flavons using modular symmetry and, simultaneously, to reproduce all the neutrino observables through the VEV of a single parameter $\tau,$ for a particular range of Yukawa coefficient factor $g.$ Two free model parameters, $\tau$ and $g$, are scanned randomly in a particular domain, and the active neutrino mass matrix is numerically diagonalized.

We have conducted the numerical analysis using the 3$\sigma$ bounds of neutrino observables so that all the neutrino observables evaluated from the model simultaneously satisfy these bounds. Our analysis of neutrino masses is also consistent with the cosmological upper bound on the sum of neutrino masses $\sum m_i <0.12$ eV.  The mixing between active and sterile neutrinos is also discussed based on the results provided by MiniBooNE, LSND and other experiments. Effective mixing angles calculated from the model are still allowed by the exclusion regions of the global analysis from $\nu_e$ and $\nu_{\mu}$  appearance and disappearance searches.

The effects of an eV-scale sterile neutrino on the active neutrino mixing angles, effective mass parameters $m_{\beta},\ m_{\beta\beta}$, and unitarity of the active neutrino mixing matrix are studied. Neutrino mixing angles are evaluated from the $(4\times 4)$ active-sterile mixing matrix $V$ by including the non-unitarity effects of sterile neutrino. Jarlskog invariant in the 3+1 sector is also determined. CP-violating Dirac phase is successfully predicted in the ranges $\delta_{CP} \sim (180.36^o - 359.73^o)$  after re-phasing the angles in the third quadrant. 

From the analysis of $m_{\beta\beta}$ and $m_{\beta}$, it is observed that the inclusion of mixing with sterile neutrino enhances the results within the future sensitivities of various experiments such as Project 8, KATRIN, Legend-1K, nEXO, etc. If these experiments fail to detect any signal, the existence of eV scale sterile neutrino will be disfavoured. The present model is also consistent with the latest upper bound of $m_{\beta\beta}$ given by KamLAND Zen in both the normal and inverted hierarchies. 

Finally, we have used the minimum $\chi^2$ analysis to predict the best-fit values of the model parameters and the neutrino oscillation data. The best-fit analysis of the present model predicts a higher octant of $\theta_{23}$. This result agrees with the latest results from the improved NOVA experiment \cite{acero2022improved}. The MES mechanism has an advantage in generating KeV-MeV scale sterile neutrino along with the active neutrino mass in the eV scale. The role of the singlet scalar $\zeta$ is to fix the scale of sterile neutrino mass and keep the term invariant under modular $A_4$. A study on the possibility of KeV-MeV sterile neutrino as a dark matter candidate will be addressed in the future. We have successfully removed any ad-hoc triplet scalars in this work and the number of free parameters is significantly reduced. In conclusion, the modular $A_4$ symmetry successfully reproduces neutrino phenomenology and other issues beyond the Standard Model without needing extra flavons as in conventional discrete symmetry models.

\section*{Acknowledgements}
We would like to thank V.V. Vien, Dept. of Physics, Tay Nguyen University, Vietnam, for the useful suggestions and comments. One of the authors(MKS) would like to thank DST-INSPIRE, govt. of India for providing the fellowship for the research under INSPIRE fellowship (ID IF180349).

\bibliographystyle{ws-ijmpa}
\bibliography{references}

\end{document}